\documentclass[a4paper,11pt]{article}
\pdfoutput=1
\usepackage{jcappub}
\usepackage[T1]{fontenc}

\usepackage{graphicx}
\usepackage{amsmath,amssymb}
\usepackage{hyperref}
\usepackage{graphicx}
\usepackage{subfigure}
\usepackage{arydshln}

\usepackage{colordvi}
\usepackage{color}

\title{\boldmath Hybrid loop quantum cosmology and predictions for the cosmic microwave background}

\author[a]{Laura Castell\'o Gomar,}
\author[a,b]{Daniel Mart\'{\i}n de Blas,}
\author[a]{Guillermo  A. Mena Marug\'an,}

\author[c,d]{and Javier Olmedo}

\affiliation[a]{Instituto de Estructura de la Materia, IEM-CSIC, Serrano 121, 28006 Madrid, Spain}
\affiliation[b]{Instituto de F\'isica, Facultad de F\'isica, 
Pontificia Universidad Cat\'olica de Chile, Av. Vicu\~na Mackenna 4860, Santiago, Chile}
\affiliation[c]{Department of Physics and Astronomy, Louisiana State University,
Baton Rouge, LA 70803-4001, USA}
\affiliation[d]{Institute for Gravitation and the Cosmos \& Physics Department, The Pennsylvania State University, University Park, PA 16802, USA}

\emailAdd{laura.castello@iem.cfmac.csic.es}
\emailAdd{damartind@uc.cl}
\emailAdd{mena@iem.cfmac.csic.es}
\emailAdd{jao44@psu.edu}

\abstract{
We investigate the consequences of the hybrid quantization approach for primordial perturbations in loop quantum cosmology, obtaining predictions for the cosmic microwave background and comparing them with data collected by the Planck mission. In this work, we complete previous studies about the scalar perturbations and incorporate tensor modes. We compute their power spectrum for a variety of vacuum states. We then analyze the tensor-to-scalar ratio and the consistency relation between this quantity and the spectral index of the tensor power spectrum. We also compute the temperature-temperature, electric-electric, temperature-electric, and magnetic-magnetic correlation functions. Finally, we discuss the effects of the quantum geometry in these correlation functions and confront them with observations.}

\begin{document}

\maketitle
\flushbottom

\section{Introduction}

In recent years, observational cosmology has experienced remarkable developments, with a considerable improvement in the resolution of the measurements \cite{planck,planck-inf}. This has provided us with a clearer view of the physics of the early Universe and with invaluable tools to investigate it. In this challenge, inflationary scenarios have played a prominent role. The inflationary paradigm has the virtue of combining simplicity with efficiency in solving several conceptual problems in cosmology, e.g. the flatness and horizon problems. Moreover, it supplies us with a mechanism capable to generate the seeds that created the large scale structures which we observe today \cite{liddle}. In this sense, cosmological perturbation theory is crucial in our present way to understand the origin of those structures and the temperature fluctuations in the cosmic microwave background (CMB). The fluctuations of the quantum geometry, which are assumed to be in fact the seeds of inhomogeneity, are described by means of linearized Einstein's theory within the framework of quantum field theory in a curved spacetime. In this context of inflationary cosmology, it suffices to give suitable initial conditions for the perturbations at the onset of inflation to reproduce, with a great deal of accuracy, the spectrum of anisotropies observed in the CMB. 

Certainly, there exist alternatives to inflation, for instance some matter bounce models \cite{matt-bounce}. They include exotic matter content that, for certain solutions, may cause a bounce in the evolution of the Universe, instead of a collapse into a singularity. Actually, other different and more generic bouncing scenarios are presently under consideration as natural mechanisms to remove the traditional big bang singularity, rather than as substitutes of inflation. The most successful one is given by loop quantum cosmology(LQC) \cite{revLQC1,revLQC2}. It is based on the quantization program of loop quantum gravity: a nonperturbative, background independent, and canonical quantization of general relativity \cite{thiem}. For homogeneous and isotropic spacetimes, and different matter contents, a neat understanding of the mechanism that cures the singularity has been reached: a quantum bounce occurs \cite{aps}. It is due to quantum geometry corrections, that dominate near the Planck regime and are able to stop the contraction of the Universe, which instead extends its evolution into an expanding branch. This regular behavior and the good control of the theoretical framework make of this formalism an appropriate arena to test quantum gravity phenomena in cosmology. 

In the last years, several approaches have been suggested to introduce small inhomogeneities in LQC, following the ideas of cosmological perturbation theory \cite{anom-free,ed,rainbow,AAN1,hybr-inf1}. These approaches try to explore the way in which quantum gravity affects the effective equations of the perturbations, with the hope that the comparison with observations will eventually permit to falsify the predictions about those quantum corrections \cite{anom-obs}. In this work, we will focus on the hybrid quantization approach \cite{hybr-inf1,hybr-rev}. It is based on the assumption that there exists a regime in which the main corrections caused by the loop quantum geometry appear in the homogeneous sector of the model, while the inhomogeneities (including those of a geometrical nature) can be treated using a standard Fock representation. This approach was proposed for the first time in Ref. \cite{hybrid}, for a linearly polarized Gowdy model with $T^3$ topology. It was proven to provide a consistent quantization of this Gowdy model \cite{hybr-rev,hybr-gow}, even in presence of matter degrees of freedom \cite{hybr-gow-matt}. Besides, in the homogeneous sector, one can restrict the study to quantum states with an effective behavior identical to that of isotropic cosmologies for certain physical properties \cite{LRS-gowdy}. It is remarkable that, for some of these states, the geometry fluctuations can then behave collectively as perfect fluids \cite{hybr-gow-FRW1,hybr-gow-FRW2}. 

The hybrid quantization approach applied to cosmological perturbations is a well developed and well understood formalism in LQC \cite{hybr-rev}. A preliminary comparison of its consequences with observations was recently presented in Ref. \cite{hybr-pred}. The approach has been implemented in full detail for Friedmann-Robertson-Walker (FRW) models with positive curvature \cite{hybr-inf1}, as well as for models with (compact) flat topology \cite{hybr-inf2,hybr-inf3,hybr-inf4}. It is possible to provide (at least formally) a complete quantization of the model, incorporating perturbations \cite{hybr-inf1}. Besides, uniqueness criteria regarding the Fock quantization of the perturbations have been put forward in this context \cite{uniq-pert}, a result that gives considerable robustness to the corresponding predictions. Concerning its treatment as a constrained theory, the system admits a first-class algebra (free of anomalies) at the quantum level \cite{hybr-inf4}. Moreover, the predictions extracted so far for the CMB turn out to be in good agreement with observations \cite{hybr-pred}. 

The analysis carried out in Ref. \cite{hybr-pred} can be considered incomplete inasmuch as it focused just on scalar perturbations and on the extraction of the temperature-temperature ($TT$) correlation function of the CMB. In the present work, we will complete the analysis by incorporating tensor modes, following the study of Ref. \cite{hybr-ten}. The inclusion of tensor perturbations is crucial to carry out an accurate comparison with observations, since single-field inflationary models generate a significant amount of these perturbations. Furthermore, it is well known that tensor perturbations provide really valuable information about the primordial stages of the Universe because, on the one hand, the effect of tensor perturbations in the TT anisotropy spectrum is more sensitive to features in the primordial power spectrum than in the case of scalar perturbations  \cite{planck-inf}  and, on the other hand, tensor perturbations are the only primordial source for magnetic-magnetic anisotropies. In order to compute the primordial power spectrum of the tensor perturbations, we will consider several adiabatic vacuum states, as well as the so-called {\it non-oscillating} vacuum state, proposed in Ref. \cite{hybr-pred}. Remarkably, we will see that a comparison of the spectra for these vacua by means of numerical techniques indicates that the non-oscillating vacuum indeed belongs to the (unitary) equivalence class of adiabatic states. Moreover, we will show that it has the asymptotic behavior of an adiabatic state of high order in the ultraviolet region of large wavenumbers. Then, supposing that the scalar and the tensor perturbations are initially in the same vacuum, we will compute the tensor-to-scalar ratio $r$. In this way we will be able to check the validity of the consistency relation  $r \simeq -8n_t$, widely used in standard  single-field inflation, where it is deduced assuming that the scalar and the tensor perturbations are both in the Bunch-Davies state \cite{allen,allen-folacci} at the onset of inflation. In addition, restricting our attention just to the non-oscillating vacuum, we will compute the electric-electric ($EE$), magnetic-magnetic ($BB$), and temperature-electric ($TE$) correlation functions. We will discuss the results, comparing them with the most recent observations of the CMB obtained by the Planck mission.

The rest of the paper is organized as follows. In Sec. \ref{sec:class} we will present the classical system. We will explain the effective dynamics of the background variables in Sec. \ref{sec:eff-dyn}, and that corresponding to the perturbations within our hybrid approach in Sec. \ref{sec:sem-dyn-hyb}. In Sec. \ref{sec:init-state} we will introduce the initial states that will be considered for the inhomogeneities. We will compute the relevant cosmological observables in Sec. \ref{sec:PS-ATPS}. Finally, we will discuss the results and conclude in Sec. \ref{sec:dis-conc}.

\section{Classical model}\label{sec:class}

Let us consider a single-field inflationary cosmological model in which the spatial sections are flat, homogeneous, isotropic, and have compact topology, isomorphic to the three-torus. The scalar field, $\Phi$, that plays the role of an inflaton, is subject to a potential $V(\Phi)$. Although the analysis that we are going to carry out is valid for quite general choices of the potential, for concreteness it will be convenient to restrict our attention to a quadratic potential of the form $V(\Phi)=m^2\Phi^2/2$. The spacetime metric is characterized by a homogeneous lapse $N_0(t)$ and by a variable $\alpha(t)$. Up to a constant, the latter is the logarithm of the scale factor, $a(t)$, that appears multiplying the auxiliary three-metric of the three-torus, $^0h_{ij}$, on each spatial section. We take spatial coordinates $\theta_i$ on these sections such that $2\pi \theta_i \in S^1$. It is useful to introduce the connection $^{0}\nabla_i$ of the static auxiliary metric $^0h_{ij}$ and the corresponding Laplace-Beltrami operator $ {^0h}^{ij} \,{^{0}\nabla_i} {^{0}\nabla_j}$. Associated with this operator, we have at our disposal the set of its real eigenfunctions, denoted by $\tilde Q_{\vec n,\epsilon} (\vec\theta)$, such that they are odd ($\epsilon=-1$) or even ($\epsilon=1$) under the transformation $\theta_i\to 1-\theta_i$. We choose these eigenfunctions to have unit norm with respect to the auxiliary volume element. Their respective eigenvalues are $-\omega_n^2=-4\pi^2\vec n\cdot\vec n$, where $\vec n=(n_1,n_2,n_3)\in\mathbb Z^3$ is any tuple in which the first nonvanishing component is a positive integer (and where, for simplicity, we obviate the zero mode). With this set of eigenfunctions, together with the connection $^{0}\nabla_i$ and the metric $^0h_{ij}$, it is possible to construct a complete basis of scalar, vector, and tensor harmonics for the spatial sections (up to the mentioned zero mode). See, e.g., Ref. \cite{hybr-inf2} for additional details. 

Around the studied homogeneous and isotropic geometries, we can now incorporate small perturbations to second order in the standard Einstein-Hilbert action, and expand them in modes using the harmonics introduced above. The scalar perturbations were already studied in Refs. \cite{hybr-inf1,hybr-inf2,hybr-inf3,hybr-inf4,hybr-pred}. The vector modes, to this level of truncation and in our model with only a scalar field, do not play any physical role, and will be ignored in the following. Therefore, we will mainly focus here on tensor perturbations. These can be described in terms of real tensor harmonics $\tilde G_{ij}$, which are eigenfunctions of the Laplace-Beltrami operator that are transverse $^0h^{ij}\;^{0}\nabla_i \tilde G_{jk}\,=0$, and traceless $^0h^{ij}\tilde G_{ij}=0$. Here, the subscript $\vec{\bf n} =(\vec n, \epsilon,\tilde{\epsilon})$ indicates the tuple $\vec n$, the parity $\epsilon=\pm$, and the polarization $\tilde \epsilon = (+,\times)$ of the tensor mode. For additional details about their definition, see e.g. the appendix of Ref. \cite{hybr-ten}. The Hamiltonian resulting from the truncation of the action to second order in the perturbations (or, strictly speaking, its zero mode, which is the only relevant part for our discussion) can be written as
\begin{equation}\label{eq:Hamiltonian}
H = N_0\Big[H_{|0}+\sum_{\vec{\bf n}}\;{^{T}\!H^{\vec{\bf n}}_{|2}}+ ({\rm scalar\;perturbations})\Big],
\end{equation} 
where the first term within the square brackets depends on the homogeneous variables only. It reads
\begin{equation}\label{eq:H_0}
H_{|0} = \frac{e^{-3\alpha}}{2}\Big(\pi_\varphi^2-{\cal H}_0^{(2)}\Big),
\end{equation}
where we have defined
\begin{align}\label{eq:H0_2}
	{\cal H}_0^{(2)}= \pi_{\alpha}^2-2 e^{6\alpha} {\bar V} (\varphi)  .  
\end{align}
The second term in Eq. \eqref{eq:Hamiltonian} is quadratic in the tensor perturbations:
\begin{align}\label{eq:TH_2}
{^{T}\!H^{\vec {\bf n}}_{|2}} &= \frac{1}{2}e^{-3\alpha}\left[\pi_{d_{\vec {\bf n}}}^2+8\pi_\alpha d_{\vec {\bf n}}\pi_{d_{\vec {\bf n}}}+2\left(5{\cal H}_0^{(2)}+3\pi_\varphi^2+4e^{6\alpha} {\bar V}(\varphi) \right)d_{\vec {\bf n}}^2+e^{4\alpha}\omega_n^2d_{\vec {\bf n}}^2\right].
\end{align}
We have called ${\bar V}(\varphi)=\sigma^4 V(\varphi/\sigma)$, with $\sigma^2=4\pi G/3$ and $G$ the Newton constant. Besides, $\pi_\alpha$, $\pi_\varphi$, and $\pi_{d_{\vec {\bf n}}}$ are the momenta conjugate to the respective variables $\alpha$, $\varphi$ (the zero mode of the scalar field, up to a constant factor \cite{hybr-inf3}), and $d_{\vec {\bf n}}$ (the variables that describe the expansion of the tensor perturbations in modes). Regarding the contribution of the scalar perturbations to the Hamiltonian, we encourage the reader to consult Refs. \cite{hybr-inf3,hybr-inf4}. 

The classical equations of motion can be easily computed by taking Poisson brackets with the total Hamiltonian \eqref{eq:Hamiltonian}, i.e.,
\begin{equation}
\dot f=\{f,H\},
\end{equation} 
where $f$ represents any function on the phase space of the system.

\section{The homogeneous sector and its effective dynamics in loop quantum cosmology}\label{sec:eff-dyn}

We now proceed to  quantize the model introduced in the previous section. As we have already mentioned, we will adopt a hybrid approach in this quantization. This means that we will combine different types of quantum representations for the different degrees of freedom of our model. In this section, we will explain the LQC quantization of the homogeneous modes and the resulting effective dynamics in this homogeneous sector. The effective dynamics of the perturbations will be explained in Sec. \ref{sec:sem-dyn-hyb}. 

Following the ideas of Refs. \cite{revLQC1,revLQC2}, one starts by adopting a description of the classical geometry in terms of an su(2)-connection and a densitized triad. On homogeneous cosmologies, they are determined by two homogeneous functions, $c$ and $p$, which respectively capture the degrees of freedom of the connection and the triad. In terms of our original phase space variables, they are given by 
\begin{equation}\label{eq:geom-new}
p = \sigma^2e^{2\alpha},\qquad pc = -\gamma \sigma^2\pi_{\alpha},
\end{equation}
where $\gamma$ is the Immirzi parameter \cite{thiem}. Here, as it is customary in LQC, we fix this parameter equal to $\gamma\simeq0{.}2375$, which is the value that allows us to recover the Bekenstein-Hawking formula in loop quantum gravity, as the leading term for large horizon area in the black hole entropy computation \cite{meissner}. 

In the so-called polymeric representation that is commonly employed in LQC, there is no well-defined operator corresponding to the connection, but rather to the holonomies of the connection. In the implementation of this polymeric representation, we will adhere to the improved dynamics scheme \cite{aps}, since this choice provides quantum geometries for which the singularity is replaced by a quantum bounce when the energy density $\rho$ reaches a constant critical value $\rho_c=3/(16 \pi G\gamma^2\Delta)\sim\, 0.41 \rho_{\rm Pl}$. Here, $\rho_{\rm Pl}$ is the Planck energy density and $\Delta$ is twice the minimum nonzero eigenvalue of the area operator in loop quantum gravity \cite{thiem}. Explicitly, $\Delta=4\sqrt{3}\pi\gamma \ell_{\rm Pl}^2$, where $\ell_{\rm Pl}$ is the Planck length. From now on, we will set up the Newton constant $G$, the reduced Planck constant $\hbar$, and the speed of light all equal to one, and work in Planck units. 

In the improved dynamics scheme, it is most convenient to pass from the triad and the connection variables $p$ and $c$ to the volume, $v=p^{3/2}$, and its conjugate variable, $\beta=c/p^{1/2}$, adopting the latter couple as basic variables. The new basic Poisson bracket is $\{\beta,v\}=4\pi \gamma$. The variable $\beta$ has a natural interpretation in the classical system: up to a constant, it is the Hubble parameter. 

On the other hand, for the homogeneous mode of the scalar field and its momentum, the usual variables employed in LQC are 
\begin{equation}\label{eq:matt-lqc}
\phi = \frac{\varphi}{\sigma},\quad \pi_\phi = \sigma\pi_{\varphi}.
\end{equation}
For convenience, we will also redefine the zero mode of the lapse function as $N=\sigma N_0$.

We now represent the homogeneous sector of the geometry on a Hilbert space ${\mathcal H}_{\mathrm{kin}}^{\mathrm{grav}}$ where the operator $\hat v$ acts by multiplication. As a distinctive property of the polymeric representation adopted in LQC, this Hilbert space admits a basis of eigenstates $\{ |\nu \rangle , \nu\in \mathbb{R}\}$ of $\hat v$ that are normalized with respect to the discrete inner product $\langle \nu_1 | \nu_2 \rangle = \delta_{\nu_1,\nu_2}$. Their eigenvalues are 
\begin{equation}
\hat v |\nu\rangle =2 \pi \gamma  \sqrt{\Delta}\, \nu |\nu\rangle.
\end{equation}
Together with this volume operator, we also have the matrix elements of the holonomies of the connection along straight edges with auxiliary length equal to $\bar \mu$, where $\bar \mu=\sqrt{\Delta/p}$ according to the improved dynamics scheme. Essentially, these matrix elements can be obtained from the operators $\hat{\mathcal{N}}_{\pm\bar\mu}$, which act on the basis states shifting their eigenvalues:
\begin{equation}\label{Nmu}
\hat{\mathcal{N}}_{\pm\bar\mu}|\nu\rangle = |\nu\pm1\rangle.
\end{equation}

For the homogeneous sector of the scalar field, we will adopt the standard representation on the kinematical Hilbert space $\mathcal H_\mathrm{kin}^\mathrm{matt}=L^2(\mathbb{R},d\phi)$, i.e., the space of square integrable functions on $\phi$ with the standard Lebesgue measure. In this representation, $\hat\phi$ acts by multiplication and $\hat\pi_{\phi}=-i\partial_\phi$. 

With this representation of the homogeneous variables, we can construct the quantum counterpart of the zero mode of the scalar constraint. In order to do so, we follow the quantization prescription of Ref. \cite{mmo}, already used in the hybrid quantization of cosmological perturbations in LQC discussed in Refs. \cite{hybr-inf1,hybr-inf2,hybr-inf3,hybr-inf4,hybr-pred,hybr-ten}. The corresponding Hamiltonian operator is defined as 
\begin{equation}\label{eq:qH_0}
\hat H_{|0} = \frac{\sigma}{2}\widehat{\left[\frac1{v}\right]}^{1/2}\hat{\mathcal C}_0\widehat{\left[\frac1{v}\right]}^{1/2},
\end{equation}
where $\hat{\mathcal C}_0$ is an operator representing the densitized version of the homogeneous part of the Hamiltonian constraint. It has the form
\begin{align}
\label{eq:calC_0}
\hat{\mathcal C}_0 &= {\hat \pi}_\phi^2- \frac{4 \pi }{3}\hat{\mathcal H}_0^{(2)}, \\	
\hat{\mathcal H}_0^{(2)} &=  \frac{3}{4 \pi }\left(\frac{3}{4 \pi \gamma^2}\hat\Omega_0^2-2 \hat v^2 V(\hat\phi)\right).
\end{align}
Here, $\hat{\Omega}^{2}_{0}$ is an operator representation of $(cp)^{2}$ in LQC. It is defined as the square of
\begin{align}\label{eq:Omega}
&\hat\Omega_0 = \frac1{4i\sqrt\Delta}\hat v^{1/2}\big[\widehat{{\rm sgn}(v)}\big(\hat{\mathcal{N}}_{2\bar\mu}-\hat{\mathcal{N}}_{-2\bar\mu}\big)+\big(\hat{\mathcal{N}}_{2\bar\mu}-\hat{\mathcal{N}}_{-2\bar\mu}\big)\widehat{{\rm sgn}(v)}\big]\hat v^{1/2},
\end{align}
where $\widehat{{\rm sgn}}$ is the sign operator and $\hat{\mathcal{N}}_{2\bar\mu}$ shifts the label of the basis states in two units [see Eq. \eqref{Nmu}].
	
The quantum states $\Psi(v,\phi)$ for the massless scalar field ($m=0$) were studied in Ref. \cite{comp}. Here, the scalar field can be regarded as a natural time function. Moreover, for highly peaked states, the evolution of the expectation values of the fundamental operators (and presumably of their products) turn out to follow the trajectories of an effective classical Hamiltonian with a high level of accuracy \cite{eff-lqc}. In this effective dynamics for LQC, the solutions depart from those of general relativity only when the energy density is at least a few percents of its critical value $\rho_c$, and in particular they avoid the big bang singularity.  
	
In the case of a massive scalar field, one has a nonzero potential $V(\hat\phi)$ in the quantum constraint. The full quantum dynamics of this system has not been studied in detail. A recent analysis in Ref. \cite{cmm} shows how one can carry out a perturbative treatment at the quantum level, valid in those situations where the contribution of the potential is small compared with the kinetic term. In the present work, nonetheless, we will consider regimes where the energy density of the scalar field is so highly dominated by its kinetic contribution in the vicinity of the bounce, that we can confidently ignore the influence of the field potential in the regions where there may be departures from general relativity. In those kinetically dominated regions, for states with large values of the scalar field momentum, all the numerical and analytic studies carried out so far in LQC strongly support the validity of the effective dynamics of LQC, as we have already mentioned, and hence we can ignore any possible further quantum contribution to the evolution of the relevant expectation values. Moreover, it is well known that in this kind of effective inflationary solutions, one can disregard all corrections arising from the regularization of the inverse-volume operator $\widehat{[1/v]}$, which is defined as
\begin{align}
&\widehat{\left[\frac1{v}\right]}^{1/3}=\frac{3}{4\pi\gamma\sqrt{\Delta}}\widehat{{\rm sgn}(v)}{|\hat v|}^{1/3}\Big(\hat{\mathcal{N}}_{-\bar\mu}{|\hat v|}^{1/3}\hat{\mathcal{N}}_{\bar\mu}-\hat{\mathcal{N}}_{\bar\mu}{|\hat v|}^{1/3}\hat{\mathcal{N}}_{-\bar\mu}\Big).
\end{align}
It has been shown that these corrections are negligible for highly peaked states, at least in the sector of states with large momentum of the scalar field (see, e.g., the discussion in Refs. \cite{sinwil,carloedward}).

In this way, we arrive at the following effective set of equations for the evolution of the expectation values of the basic operators of the homogeneous sector of our model:
\begin{subequations}\label{eq:hom-eqs2}
\begin{align}
\frac{1}{N}\dot{\phi} &=  \frac{\pi_\phi}{v},\\\label{eq:dot-piphi2}
\frac{1}{N}\dot{\pi}_\phi &= -v\frac{dV(\phi)}{d\phi},
\\\label{eq:dot-vol2}
\frac{1}{N}\dot v &= \frac{3}{2} v\frac{\sin(2\sqrt{\Delta}\beta)}{\sqrt{\Delta}\gamma},\\
\frac{1}{N} \dot \beta &= -\frac{3}{2} \frac{\sin^2(\sqrt{\Delta}\beta)}{\Delta\gamma}+4\pi\gamma \left(V(\phi)-\frac{\pi_\phi^2}{2v^2}\right).
\end{align}
\end{subequations}
The dot denotes the time derivative with respect to an arbitrary time function $t$. In addition, the effective homogeneous Hamiltonian can be written in the form $\pi_{\phi}^2=4\pi {\mathcal H}_0^{(2)}/3$, where
\begin{align}	\label{eq:H_02}
{\mathcal H}_0^{(2)} &=  \frac{3}{4 \pi}\left(\frac{3}{4\pi \gamma^2}\frac{v^2\sin^2(\sqrt{\Delta}\beta)}{\Delta}-2  v^2 V(\phi)\right).
\end{align}
Here, we have neglected backreaction contributions coming from the inhomogeneities. Therefore, the effective equations of motion of the background coincide with the usual ones in LQC \cite{as}. Also, in order to simplify the notation, we have dropped the expectation value symbols.

It is worth remarking that, in this homogeneous model, any effective trajectory can be determined by the value of the scalar field at the bounce, $\phi_B$. Actually, the value of the volume, $v_B$, (or, equivalently, of the scale factor) at the bounce, can be reset arbitrarily, e.g. fixing it equal to one, because there is no intrinsic absolute length scale, owing to the homogeneity and the lack of spatial curvature. Besides, the time derivative of this volume vanishes at the bounce. Finally, the momentum conjugate to the scalar field is determined by the Hamiltonian constraint. Therefore, for a fixed value of the mass $m$ of the inflaton field and a given choice of the Immirzi parameter $\gamma$, a single piece of data turns out to specify the solutions of our homogeneous model. Furthermore, since the energy density is bounded from above in LQC, with the bound reached at the moment of the bounce, we conclude that, in the considered case of the quadratic potential the mass of the inflaton field and the value of the scalar field at the bounce must satisfy the inequality
\begin{equation}
 m^{2}\phi_{B}^2 \leq 0.82. 
\end{equation}

\section{Effective dynamics of the perturbations in the hybrid approach}\label{sec:sem-dyn-hyb}

In order to complete the quantization of the full system, including the inhomogeneities, and deduce effective equations of motion for the perturbations, we will introduce a Fock representation for them and define their quadratic contribution to the quantum Hamiltonian constraint.  Actually, the Fock representation of the scalar perturbations was detailed in Ref. \cite{hybr-inf3}. Thus, here we will focus on the quantization of the tensor perturbations. For these tensor modes, the hybrid quantization approach was implemented in Ref. \cite{hybr-ten}. Let us summarize the more important steps. Previous to the quantization, one performs the canonical transformation
\begin{align}\label{eq:can-trans}
\tilde d_{\vec {\bf n}}=e^{\alpha}  d_{\vec {\bf n}},\quad \pi_{\tilde d_{\vec {\bf n}}}=e^{- \alpha}\big(\pi_{ d_{\vec {\bf n}}}+3\pi_{\tilde \alpha} d_{\vec {\bf n}}\big).
\end{align}
This transformation must be extended to the homogeneous sector by including appropriate quadratic contributions of the perturbations in the definition of the canonical variables for the homogeneous geometry \cite{hybr-ten}. However, we will ignore these corrections in this article, admitting that they are sufficiently small (in fact, they can be interpreted as a kind of backreaction correction to the definition of the background variables). The resulting (zero mode of the) Hamiltonian constraint is still of the form \eqref{eq:Hamiltonian}, but with the quadratic tensor term replaced  with
\begin{align}\label{eq:tildeTH_2}
{^{T}\!{\tilde H}^{\vec {\bf n}}_{|2}} &= \frac{1}{2}e^{- \alpha}\left[{\pi}_{\tilde d_{\vec {\bf n}}}^2+\left(e^{-4 \alpha}{\cal H}_0^{(2)}-4 e^{2\alpha} {\bar V} (\varphi)+\omega_n^2\right){\tilde d}_{\vec {\bf n}}^2\right].
\end{align}
Hence, it follows that the evolution of the tensor modes, at the classical level, is given by a set of second-order linear differential equations, involving time-dependent coefficients. 

To incorporate the perturbations in the quantum theory, one adopts the Fock representation introduced in Ref. \cite{hybr-ten} for the tensor modes. For the subsequent representation of the quadratic perturbative contribution \eqref{eq:tildeTH_2}, one casts its background dependence in terms of the variables used in LQC, given by Eqs. \eqref{eq:geom-new} and \eqref{eq:matt-lqc}. Then, the considered Hamiltonian constraint becomes
\begin{equation}\label{eq:C_0}
\hat H = \frac{\sigma}{2}\widehat{\left[\frac1{v}\right]}^{1/2}\left[\hat{\mathcal C}_0-\hat{\Theta}^T+({\rm scalar\;perturbations})\right]\widehat{\left[\frac1{v}\right]}^{1/2}.
\end{equation}
The term associated with the tensor perturbations reads
\begin{equation}
\hat{\Theta}^T = -\sum_{\vec {\bf n}}\left[\left(\hat{\vartheta}\,\omega_n^2+\hat{\vartheta}_T^q \right)\hat{\tilde d}_{\vec {\bf n}}^2 + \hat{\vartheta}\, \hat{\pi}_{\tilde d_{\vec {\bf n}}}^2\right].
\end{equation}
The $\vartheta$-operators are functions exclusively of other operators that have already been defined in the representation of the homogeneous sector. Their explicit expressions are
\begin{align}
\hat \vartheta&=\hat v^{2/3},\\ \hat \vartheta_T^q&=\frac{16\pi^2}{9}\widehat{\left[\frac1{v}\right]}^{1/3}\hat{\mathcal H}_0^{(2)}\widehat{\left[\frac1{v}\right]}^{1/3}\!\!\!\!-\frac{16\pi}{3}\hat v^{4/3}V(\hat \phi).
\end{align}
In presence of inhomogeneities, the solutions to the constraint are not known explicitly, although one can always carry out a formal quantization following the ideas of Ref. \cite{hybr-inf1}. Here, we will follow instead the strategy of Refs. \cite{hybr-inf3,hybr-inf4}, where one adopts a Born-Oppenheimer ansatz for the solutions to the constraint, $\Xi$, so that there is a separate dependence on the background geometry and on the perturbations:
\begin{equation}\label{BOans}
\Xi(v,\phi,\tilde d_{\vec {\bf n}})=\Psi(v,\phi)\psi(\tilde d_{\vec {\bf n}},\phi).
\end{equation}
In this formula, we have ignored the scalar perturbations, and $\Psi(v,\phi)$ is a solution to the background homogeneous constraint ({\emph{sufficiently accurate} at the perturbative level in which one wants to allow for backreactions effects). On the other hand, $\psi(\tilde d_{\vec {\bf n}},\phi)$} is the wave function of the tensor perturbations defined on a suitable Fock space, ${\cal F}$, once the homogeneous scalar field $\phi$ is regarded as an internal time. This ansatz has already been studied in the context of tensor perturbations in Ref. \cite{hybr-ten}. Under reasonable conditions (similar to those explained for the scalar perturbations in Refs. \cite{hybr-inf3,hybr-inf4}, and expected to hold in semiclassical regimes), it is possible to deduce a Schr\"odinger equation for the perturbations, with a physical Hamiltonian that rules the evolution in the time $\phi$ given by
\begin{align}\label{eq:BOans-constGamma3}
\hat H^T_{\rm phys}=\frac{\langle\hat{\Theta}^T\rangle_\Psi}{2\big\langle{\sqrt{\hat{\mathcal H}_0^{(2)}}}\big\rangle_\Psi}.
\end{align}
These expectation values are taken over the homogeneous geometry, with the inner product of LQC. The operator in the denominator can be understood as the square root of the positive part of $\hat{\mathcal H}_0^{(2)}$. This physical Hamiltonian is a well defined operator acting on $\psi(\tilde d_{\vec {\bf n}},\phi)$. 

When the effective dynamics of LQC is valid for the background geometry, one then arrives to effective equations of motion for the tensor perturbations that, in conformal time, can be combined into the following second-order differential equation\footnote{In fact, the conditions to derive a Schr\"odinger equation are not necessary to obtain Eq. \eqref{eq:KG-eq}. Arguments like those of Refs. \cite{hybr-inf3,hybr-inf4} show that it essentially suffices that: a) one can ignore geometry transitions mediated by the (zero mode of the) Hamiltonian constraint after introducing the Born-Oppenheimer ansatz, b) the effective dynamics of LQC is valid, and c) the quadratic perturbative terms admit an effective description obtained with the direct classical counterpart of the annihilation and creation operators.} 
\begin{align}\label{eq:KG-eq}
{\tilde d}^{\prime\prime}_{\vec {\bf n}}+\left(\omega_n^2+\left(\frac{4 \pi}{3}\right)^2\frac{1}{v^{4/3}}{\mathcal H}_0^{(2)}-\frac{16\pi }{3}v^{2/3}V(\phi)\right) {\tilde d}_{\vec {\bf n}}=0.
\end{align}
Here, the prime stands for the derivative with respect to the conformal time $\eta$, or which $N = v^{1/3}$. A similar equation has been derived for the scalar perturbations in Ref. \cite{hybr-inf4}, and its physical consequences have been partially studied in Ref. \cite{hybr-pred}. As one would expect, these equations do not depend on the compactification scale of the three-torus (or, equivalently, on the period chosen for our coordinates $\theta_i$ \cite{hybr-inf4, hybr-ten}). Then, one can remove that compactification scale and pass to a continuous description in which the discrete eigenvalues $\omega_n$ become a wavenumber $k$ that can take any positive real value. In the following, we adopt this continuous formulation.

\section{Initial state of the tensor perturbations}\label{sec:init-state}

Our next task is to select a suitable initial state for the perturbations. Together with their effective equations of motion, this will determine their value in the evolution. In this way, we will be able to extract predictions about their primordial spectrum that eventually could be compared with observations. Here, we will mainly follow the traditional procedures in cosmological single-field inflation, based on quantum field theory in curved spacetimes. Let us start with the choice of an initial time, $\eta_i$. Although any arbitrary choice is possible, we will consider the bounce as a natural initial Cauchy surface to give initial conditions, and we will denote the corresponding time as $\eta_{B}$. Obviously, we are not saying that other choices of initial time are not acceptable. Other alternatives, like e.g. the limit of infinitely negative conformal time, may be worth exploring.

Since we have adopted a Fock representation for the tensor perturbations, an equivalent way to fix their initial vacuum state is to specify a complete orthonormal set of positive frequency solutions to the equations of motion of the field. We choose these solutions so that they do not mix modes (actually, this guarantees translational invariance on the spatial sections) and coincide for all modes with the same wavenumber $k$ (guaranteeing rotational symmetry in the considered continuous case). Moreover, we take such a set of complex solutions $\{\mu_k(\eta)\}$ to be orthonormal with respect to the usual Klein-Gordon inner product:
\begin{equation}\label{eq:inner-prod}
\Big(\mu_k^{(1)},\mu_k^{(2)}\Big)=i\left[\big(\mu_k^{(2)}\big)^* \mu_k^{\prime (1)}-\mu_k^{(1)}\big(\mu_k^{\prime (2)}\big)^*\right],
\end{equation}
where $i$ is the unit imaginary number and the star symbol stands for complex conjugation (in this expression, and in what follows, we do not display explicitly the time dependence of $\mu_k$, and the prime denotes again the conformal time derivative). Since the equations are of second order and possess real (time-dependent) coefficients, a complete set of linearly independent solutions is given by $\mu_k$ and its complex conjugate $\mu_k^*$, provided the former has indeed unit norm: $(\mu_k,\mu_k)=1$. Actually, one can easily check that $(\mu_k,\mu_k^*)=0$, and that $\mu_k^*$ is normalized in the sense that $(\mu_k^*,\mu_k^*)=-1$. These orthonormality conditions are fulfilled at any time, because the Klein-Gordon inner product is preserved on shell in the evolution. In particular, they must be satisfied at the initial time $\eta_i$, that we let arbitrary for the moment. These conditions constrain the freedom in the choice of initial data. In terms of our solutions, the variables rep
 resenting the perturbations then have the form
\begin{equation}
\tilde{d}_{\vec{\bf k}}(\eta) = \mu_{k}(\eta) a_{\vec{\bf k}} + \mu_{k}^{\ast}(\eta) a^{\ast}_{\vec{\bf k}}
\end{equation}
where $a_{\vec{\bf k}}$ and $a^{\ast}_{\vec{\bf k}}$ are, respectively, time-independent annihilation and creation variables for the mode $\vec{\bf k}$ (with wavenumber equal to $k$).

Summarizing, the choice of initial data $\mu_k(\eta_i)$ and $\mu_k^{\prime}(\eta_i)$ for the sector of positive frequency, orthonormalized with respect to the Klein-Gordon inner product, completely determines the initial vacuum state of the perturbations. This is indeed equivalent to introducing a complex structure \cite{wald}. We recall that a complex structure $J$ is a real linear transformation in the complex vector space of solutions such that $J^2=-1$. Besides, $J$ must be compatible with the inner product \eqref{eq:inner-prod} (in the sense that an appropriate composition of $J$ with the inner product provides a positive bilinear map \cite{wald}). Any complex structure induces a splitting of the space of solutions into two orthogonal subspaces, that are usually identified with the positive and negative frequency sectors. The freedom in the choice of complex structure $J$ is equivalent to the freedom in the choice of orthonormalized initial data for the positive frequency solutions, and therefore to the selection of an initial vacuum state of the field.

In our case the initial data, and in consequence the initial vacuum of the tensor modes, can be parameterized in terms of two real functions for each mode $k$. If we call $\mu_{k,0}=\mu_k(\eta_i)$ and $\mu^{\prime}_{k,0}=\mu^{\prime}_k(\eta_i)$, any arbitrary set of initial conditions, up to an irrelevant global phase, can be written as
\begin{equation}
\label{eq:gen-ini-cond}
\mu_{k,0} = \frac{1}{\sqrt{2D_{k}}}, \quad \mu^{\prime}_{k,0}= \sqrt{\frac{D_{k}}{2}}\left(C_{k}-i\right).
\end{equation}
We restrict the function $D_{k}$ to be strictly positive, whereas $C_{k}$ can take any real value. Although one can choose freely these mode functions, there exist natural restrictions on them based on physical arguments. These restrictions refer mainly to their ultraviolet behavior. For instance, the requirement of a unitary dynamics \cite{uniq-pert,uniq-t3,unit-dsitter} employed in the hybrid quantization that we have adopted \cite{hybr-ten}, as well as the prescriptions of adiabatic states that are typical in inflationary contexts \cite{adiab-LR,adiab-reg-PF,adiab-reg-AP}, or the Hadamard condition \cite{wald}, they all restrict the asymptotic ultraviolet behavior in the form\footnote{After the scaling of the tensor modes carried out in Eq. \eqref{eq:can-trans}, this asymptotic behavior ensures that the requirement of unitary dynamics picks out a unitary equivalence class of vacua that includes the Hadamard and the adiabatic states.} $D_{k} = k + o(k^{-1/2})$ and $C_{k} = o(k^{-3/2})$ for infinitely large $k$, where the symbol $o(k^{l})$ denotes terms that are negligible compared to $k^l$ for a given power $l$.

In this work, we will consider two different types of prescriptions for the selection of initial data. The first one is based on adiabatic states. We will consider two ways to select a specific set of initial data of 0th, 2nd, and 4th adiabatic order, following constructions that are similar to those described in Ref. \cite{adiab-LR} and in Refs. \cite{adiab-reg-PF,adiab-reg-AP}, respectively. The second prescription corresponds to the non-oscillating vacuum that was introduced in Ref. \cite{hybr-pred}. Both prescriptions will be detailed below.

\subsection{Adiabatic states}

Adiabatic states were originally introduced as approximated solutions to the equations of fields propagating in cosmological spacetimes. They are also considered in cosmology as a way to prescribe initial conditions for the quantum fields with convenient physical properties. As we will see, these states provide initial data with a suitable behavior for asymptotically large $k$. Here, for the sake of brevity, we will define them only for the particular model under consideration. 

Let us adopt the following ansatz for the solutions:
\begin{equation}
\label{eq:adiab-sol}
\mu_{k} = \frac{1}{\sqrt{2W_{k}(\eta)}}e^{-i\int^{\eta}W_{k}(\bar{\eta})d\,\bar{\eta}}.
\end{equation}
If we substitute this into Eq. \eqref{eq:KG-eq}, we obtain for $W_{k}$ the differential equation
\begin{equation}
\label{eq:adiab-sol-W}
W^{2}_{k} = k^{2} + s - \frac{1}{2}\frac{W^{\prime\prime}_{k}}{W_{k}}+\frac{3}{4}\left(\frac{W^{\prime}_{k}}{W_{k}}\right)^{2}.
\end{equation}
The function $s=s(\eta)$ is given by the time-dependent mass of the corresponding Klein-Gordon equation. In the case of the tensor perturbations in the hybrid approach, we have
\begin{equation}
s^{(t)} = \left(\frac{4 \pi}{3}\right)^2\frac{1}{v^{4/3}}{\mathcal H}_0^{(2)}-\frac{16\pi }{3}v^{2/3}V(\phi).
\end{equation}
Different adiabatic solutions $W^{(\mathfrak{n})}_{k}$, where $\mathfrak{n}$ is an integer that indicates the adiabatic order, are in fact approximations to the exact solutions $W_{k}$. Each of them converges to the exact one at least as $\mathcal{O}(k^{-1-\mathfrak{n}})$ in the limit $k\to\infty$, where the symbol $\mathcal{O}$ stands for asymptotic order\footnote{Our definition of adiabatic order $\mathfrak{n}$ differs from others in the literature \cite{AAN2} for which the convergence rate is $\mathcal{O}(k^{-\mathfrak{n}})$. Our convention is motivated by the fact that the asymptotic expansion of $W_k$ does not contain even inverse powers, and we start the counting at 0th order.}. Therefore, this method provides good approximate solutions for the ultraviolet modes. However, this is not necessarily the case for small $k$. Indeed, while adiabatic states constrain the behavior of the solutions in the asymptotic limit of large $k$, i.e., at small scales, they still allow for an infinite freedom in the behavior of the large scale solutions. Even so, this prescription has proven very useful in order to obtain analytic expressions approximating the exact solutions \cite{adiab-LR}, as well as for the renormalization of the stress-energy tensor in cosmological scenarios \cite{adiab-reg-AP,adiab-reg-PF}. 

Within LQC, adiabatic states have been used to specify initial data in inflationary models that are closely related to the one under study here, for instance in Ref. \cite{AAN2} for the dressed metric approach, and in Ref. \cite{hybr-pred} for the scalar perturbations in the context of the hybrid quantization approach. These initial data associated with adiabatic states are given in the form  \eqref{eq:gen-ini-cond} with
\begin{align}
D_{k} = W_{k},\qquad C_{k} = -\frac{W^{\prime}_{k}}{{2W^{2}_{k}}}.
\end{align}
Here, $W_{k}$ and its time derivative have to be evaluated at the chosen initial time. To get initial data for different adiabatic orders, one only has to replace $W_{k}$ in the previous expression with the adiabatic solution at the order in question.

In this article, we are going to consider the initial data for different adiabatic orders obtained by two different constructions, as it was done also in Ref. \cite{hybr-pred}. The first construction follows ideas presented in Ref. \cite{adiab-LR}. A solution of order $(\mathfrak{n} + 2)$, i.e. $W^{(\mathfrak{n}+2)}_{k}$, is obtained by plugging $W^{(\mathfrak{n})}_{k}$ in the right-hand side of Eq. \eqref{eq:adiab-sol-W}. This process is carried out iteratively, starting with $W^{(0)}_{k} = k$, the 0th-order adiabatic state. It is worth mentioning that $W^{(0)}_{k}$ corresponds to the natural solution of a free massless scalar field in a Minkowski spacetime. In the second construction, one performs an asymptotic expansion of the solution $W_{k}$ in inverse powers, in the limit $k\rightarrow\infty$, and truncates this expansion at the considered order. We will call $\mathfrak{W}^{(\mathfrak{n})}_{k}$ the functions obtained in this way. This method is analogous to the one used in Ref. \cite{AAN2}, and the resulting state is known in the literature as the {\it obvious} adiabatic state of $\mathfrak{n}$th order. 

For each of these two constructions of adiabatic states, we will consider here the adiabatic initial conditions of 0th, 2nd, and 4th order, which are determined by the functions
\begin{align}
& W^{(0)}_{k} = \mathfrak{W}^{(0)}_{k} = k, \\
& W^{(2)}_{k} = \sqrt{k^{2}+s},\qquad \mathfrak{W}^{(2)}_{k} = k + \frac{s}{2k}, \\
&  W^{(4)}_{k} = \sqrt{k^{2} + s + \frac{5}{16}\left(\frac{s^{\prime}}{k^{2} + s}\right)^2 - \frac{s^{\prime\prime}}{4(k^{2}+s)}}, \qquad \mathfrak{W}_{k}^{(4)} = k + \frac{s}{2k} - \frac{s^{2}+s^{\prime\prime}}{8k^{3}}. 
\end{align}
The derivatives of the time-dependent mass that appear in these expressions are calculated using the effective equations of motion of the homogeneous variables. 

It is obvious that the two constructions provide different initial conditions (except at 0th order). Besides, none of the two procedures can be considered rigorously sound, inasmuch as there is no guarantee that the corresponding initial conditions are meaningful for all values of $k$ and independently of the behavior of the time-dependent mass. In order to explain this statement, let us consider, for instance, the 2nd-order adiabatic solutions $W^{(2)}_{k}$ and $\mathfrak{W}^{(2)}_{k}$. If one selects an initial time in which the time-dependent mass is negative, then $W^{(2)}_{k}$ provides meaningful initial conditions only for $k > \sqrt{-s}$, whereas $\mathfrak{W}^{(2)}_{k}$ does so only for $k > \sqrt{-s/2}$. Therefore, in this situation, none of the two constructions can be trusted in order to determine a complete set of physically acceptable initial data. Fortunately, in the case of the hybrid approach (and in the regimes that we are interested to discuss), the time-dependent mass of the tensor perturbations, and also that of the scalar perturbations, turn out to be both strictly \emph{positive} at the bounce (and close to it). For completeness, let us recall that the expression of the mass of the scalar perturbations is \cite{hybr-pred},

\begin{equation}
 s^{(s)} = \frac{16\pi^{2}\mathcal{H}_{0}^{(2)}}{9v^{4/3}}\left(19 - \frac{32 \pi^{2} \gamma^{2} \mathcal{H}_{0}^{(2)}}{\Omega^{2}}\right) + v^{2/3}\left(\frac{d^{2}V(\phi)}{d\phi^{2}} + \frac{16 \pi \gamma \pi_{\phi} \Lambda }{\Omega^{2}}\frac{dV(\phi)}{d\phi} - \frac{16\pi}{3}V(\phi) \right),
\end{equation}
where $\Lambda = |v| \sin{(2\sqrt{\Delta}\beta )}/(2\sqrt{\Delta})$. This expression is equivalent to the one provided in Ref. \cite{hybr-pred}, modulo the homogeneous constraint.
Evaluating also the first and second derivatives of $s^{(t)}$ and $s^{(s)}$ at the bounce, we have checked that all the adiabatic initial conditions considered here are well defined for all modes with our hybrid approach, at least for the set of initial values of the background variables that we have explored in our numerical simulations.

At this stage of our discussion, it is also worth noting that, if one considers instead the dressed metric approach, the time-dependent mass becomes \emph{negative} when one approaches the bounce\footnote{While the effective equations of the perturbations in the hybrid approach have a smooth transition connecting the collapsing and he expanding branches, such that general relativity is recovered asymptotically in the two branches, it is not obvious how these two properties of smoothness and semiclassicality can be attained in the dressed metric approach on the union of both branches. For this reason, we will only compare the results of the two approaches in the expanding branch.}. Let us recall that the dressed metric approach and the hybrid approach are the two only proposals within the framework of LQC that lead to hyperbolic equations for the perturbations, and at present are the only proposals in this framework that can be considered compatible with observations \cite{Bolliet}. As it was already pointed out in Ref. \cite{hybr-pred}, the LQC corrections that appear in the hybrid and the dressed metric approaches differ slightly, owing to the different strategies that are followed in the quantization, a fact that leads to distinct predictions for the primordial spectra. Even without carrying out a direct comparison, it is easy to see that (when one considers in the dressed metric approach a scaling of the perturbations like the one performed in Eq. \eqref{eq:can-trans}, namely, by a factor $\tilde{a}=e^{\alpha}$), the time-dependent mass for the tensor perturbations is \cite{AAN2}
\begin{equation}
\tilde{s}^{(t)}(\eta) = -\frac{\tilde{a}^{\prime\prime}}{\tilde{a}},
\end{equation}
whereas, for the scalar perturbations in the expanding phase, one obtains
\begin{equation}
\tilde{s}^{(s)}(\eta) = -\frac{\tilde{a}^{\prime\prime}}{\tilde{a}} + \tilde{a}^{2}\left(\mathfrak{f}^{2}V(\phi)+ 2 \mathfrak{f}\,\frac{dV(\phi)}{d\phi}+\frac{d^2V(\phi)}{d\phi^2}\right).
\end{equation}
Here, $\mathfrak{f} = \sqrt{24\pi}\phi^{\prime}/\sqrt{\tilde{a}^{2}\rho}$. Note that $\tilde{a}$ is proportional to the scale factor $a=v^{1/3}$ of the effective dressed metric. At the time of the bounce, $\tilde{a}^{\prime\prime}$ is positive, and therefore the time-dependent mass is negative for the tensor perturbations (as well as for the scalar perturbations in bounces that are dominated by the kinetic energy). Therefore, one is forced to consider different constructions for the initial conditions, or initial times that render them physically meaningful.
\begin{figure}
\centering
\includegraphics[width=0.49\textwidth]{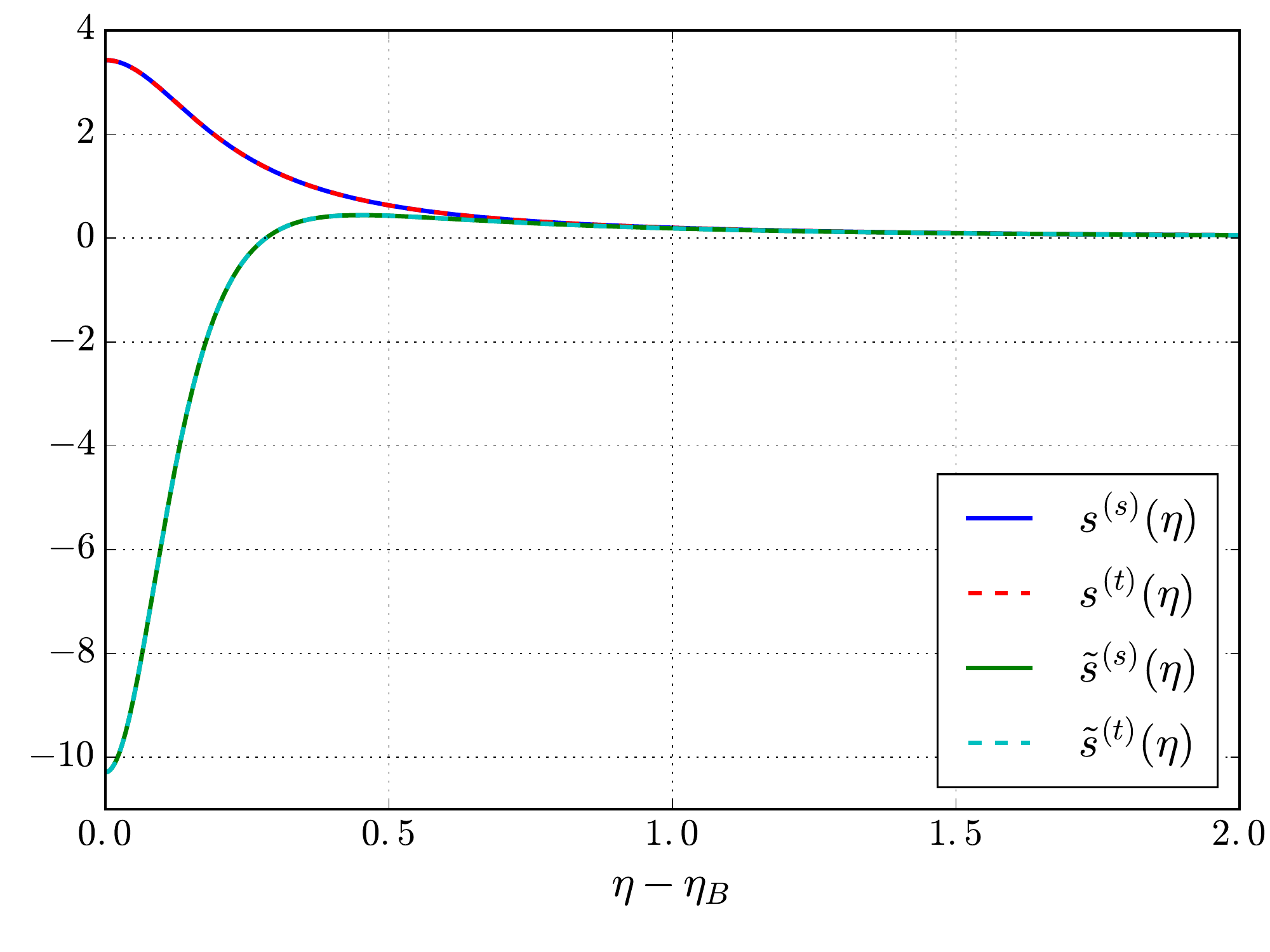} 
\includegraphics[width=0.49\textwidth]{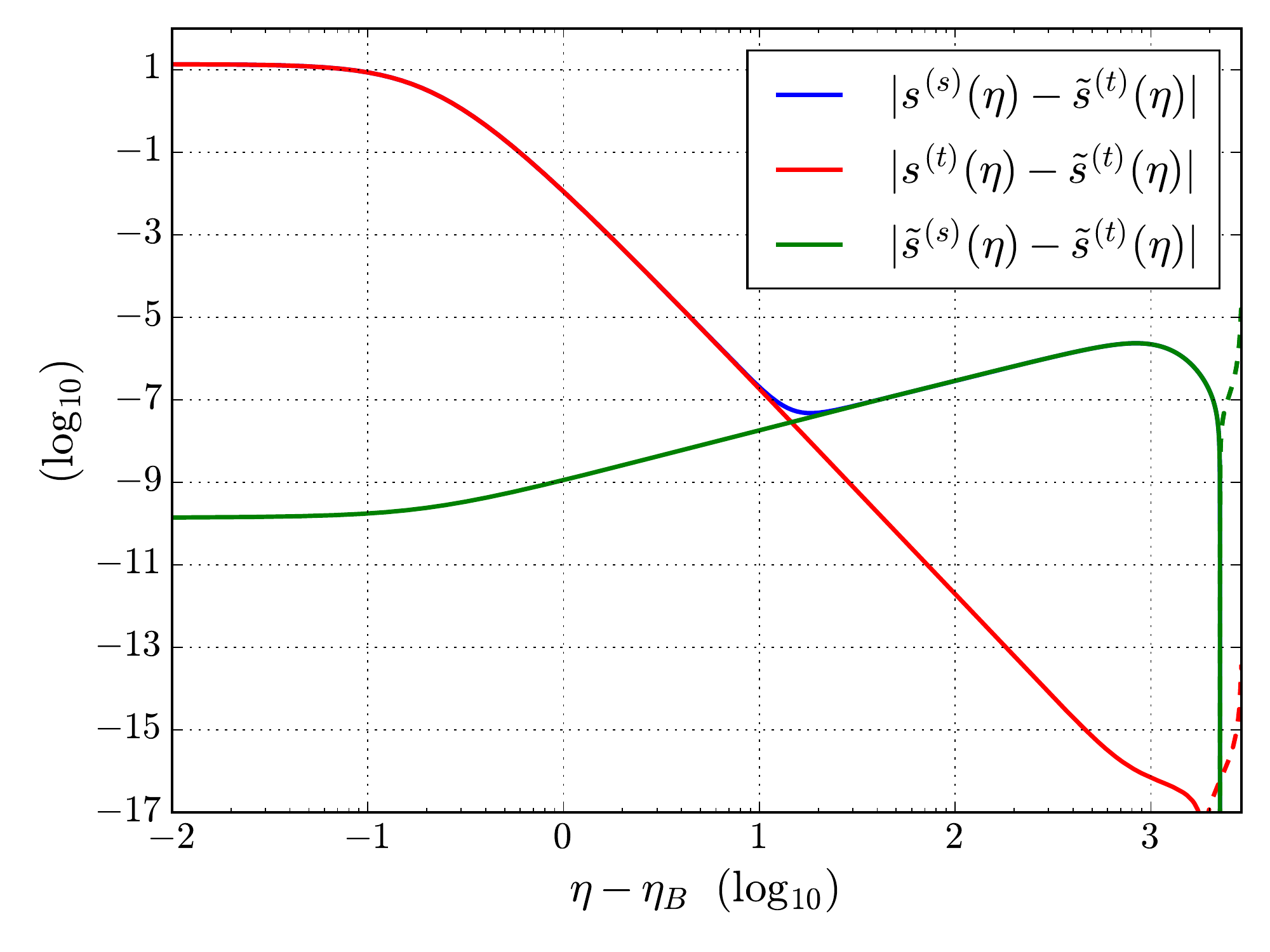} 
\caption{Comparison of the time-dependent mass of the tensor and the scalar perturbations for the hybrid approach and the dressed metric approach in a dynamical trajectory with $\phi_{B} = 0.97$ and $m = 1.20 \cdot 10^{-6}$ in Planck units (kinetically dominated bounce). Here, $s^{(s)}(\eta)$ and $s^{(t)}(\eta)$ are the time-dependent masses of the scalar and the tensor perturbations in our hybrid approach, respectively, while $\tilde s^{(s)}(\eta)$ and $\tilde s^{(t)}(\eta)$ are their counterparts in the dressed metric approach. Dashed lines in the right panel indicate negative values of the quantity of which we plot the absolute value.}
\label{fig:tdmass}
\end{figure}
In Fig. \ref{fig:tdmass} we compare the evolution of the time-dependent mass of the tensor and the scalar perturbations for the two mentioned approaches. Analyzing each of the two prescriptions separately, we observe that the values and the evolution of the time-dependent mass for the two types of perturbations are almost identical, including the region around the bounce. On the contrary, if we compare the two prescriptions, we see that the time-dependent masses differ around the bounce, where the LQC corrections are important, but then they quickly converge to the same values, far enough away from that bounce. Clearly, this shows that the effective equations of motion for the perturbations are not the same in the two approaches, even if one neglects the backreaction in both cases.

Turning back to the issue of the election of initial data for the perturbations, we emphasize that, although the adiabatic conditions that we have discussed reduce the freedom of choice, there is still an infinite number of possible adiabatic states at any order. Then, additional criteria must be required in order to remove this freedom. For instance, in Ref. \cite{pref-inst-vac-AAN} it has been proposed that one should select the state that provides a regularized stress-energy tensor which vanishes at the given initial time. Although there is an infinite ambiguity in the adiabatic renormalization process, once one fixes that ambiguity (as it is done in Ref. \cite{pref-inst-vac-AAN}) from the point of view of an observer at the end of inflation, one is left only with a one-parameter family of states that arise from the remaining freedom in the specification of the initial time. If this initial time is fixed, the vacuum state turns out to be unique.

\subsection{Non-oscillating vacuum}

As an alternative to these adiabatic considerations, Ref. \cite{hybr-pred} puts forward another criterion that seems to identify also a unique set of initial data for the perturbations. This criterion can be understood in terms of a variational problem for the data \eqref{eq:gen-ini-cond}. The coefficients $D_{k}$ and $C_{k}$ are selected so that the time variation of the power spectrum associated with the 2-point function gets minimized on an appropriate interval. In fact, this criterion, when applied to a massive (or massless) scalar field in Minkowski spacetime, or to a massless scalar field in the cosmological chart of de Sitter spacetime, picks out the Poincar\'e vacuum state, or the Bunch-Davies one, respectively. For de Sitter spacetime, hence, the criterion is equivalent to de Sitter invariance plus the Hadamard condition.

In more detail, let us consider the 2-point function of the tensor perturbations at the end of inflation. Its power spectrum (obtained from its Fourier transform) is 
\begin{equation}\label{eq:PS}
{\cal P}_{\cal T}(k) =\frac{32 k^3}{\pi}\frac{|\mu_{k}|^2}{a^{2}}\bigg|_{\eta=\eta_{\rm end}},
\end{equation}
where we recall that $a=v^{1/3}$ is the scale factor (see, for instance, Refs. \cite{liddle,lang}). The criterion to select the so-called non-oscillating vacuum is based on a specific choice of the functions $D_{k}$ and $C_{k}$ that determine the initial data of each mode solution $\mu_{k}$. The choice is such that the oscillations in the power spectrum during the evolution, oscillations that are often naively attributed to particle creation, are minimized in a given time interval: in the present situation, the interval from the bounce to the instant at which $\phi^{\prime} = 0$. In this way, the criterion selects the initial conditions that minimize the temporal variation of $|\mu_{k}|^2$ in the studied period of time. In order to determine these initial conditions, we define the quantity
\begin{equation}\label{eq:IO}
\int_{\eta_{i}}^{\eta_{f}}\left|\frac{d\big( |\mu_{k}|^{2}\big)}{d\bar{\eta}}\right|d\bar{\eta}
\end{equation}
for each mode, where $\eta_f$ is a final time. This integral depends on the initial conditions and the dynamical equations through the mode solution $\mu_{k}$, as well as on the considered interval of integration $({\eta_{i}},{\eta_{f}})$ (obviously, this implies a nonlocal dependence). To find out the desired values of $D_{k}$ and $C_k$, we vary them in order to minimize \eqref{eq:IO}. Analytic calculations are possible in some specific scenarios. For instance, the appendix of Ref. \cite{hybr-pred} contains a detailed computation for the case of a massless scalar field in a de Sitter cosmological spacetime. It turns out that, in that case, the set of initial data that minimizes the integral is unique and reproduces the Bunch-Davies vacuum provided that the (conformal) time interval under consideration starts at minus infinity. In that reference, initial data were also determined such that they minimize the temporal oscillations of the power spectrum for the scalar perturbations in LQC when the time interval goes from the bounce to the moment in which the kinetic energy of the field vanishes. This last computation was carried out using numerical techniques, since no analytic tools were available. In this sense, we would like to call the reader's attention to Ref. \cite{anal}, where approximated analytic expressions have been deduced for the perturbations in the dressed metric approach. Those expressions might be useful as well in the context discussed here, although they would have to be extended first to the hybrid approach. Finally, let us comment that the results of our computations show that the non-oscillating vacuum determined with the above numerical method does not depend significantly on the selection of the bounce as the initial instant of time in the interval of integration. This initial time can be changed in a surrounding of the bounce without affecting much the form of the non-oscillating vacuum state.

\subsection{Numerical analysis of the initial conditions}

In this subsection, we want to compare the initial conditions corresponding to adiabatic states of different orders and to the non-oscillating vacuum. In the following, we will take the bounce as initial Cauchy surface, i.e., $\eta_i=\eta_B$. With this choice, it is easy to specify initial data for the background variables, as we have already explained. Besides, the main corrections of quantum gravity nature happen in the bouncing regime. Therefore, we expect that the perturbations at the end of inflation may keep memory of the physical processes around the bounce.

We will carry out the comparison of the initial data in two different manners. First, we will make a quantitative analysis, comparing the functions $D_k$ and $C_k$ that parameterize the initial conditions and define the corresponding annihilation and creation variables, restricting the study to the set of wavenumbers $k$ that are of physical relevance in cosmology. 
\begin{figure}
\centering
\includegraphics[width=0.49\textwidth]{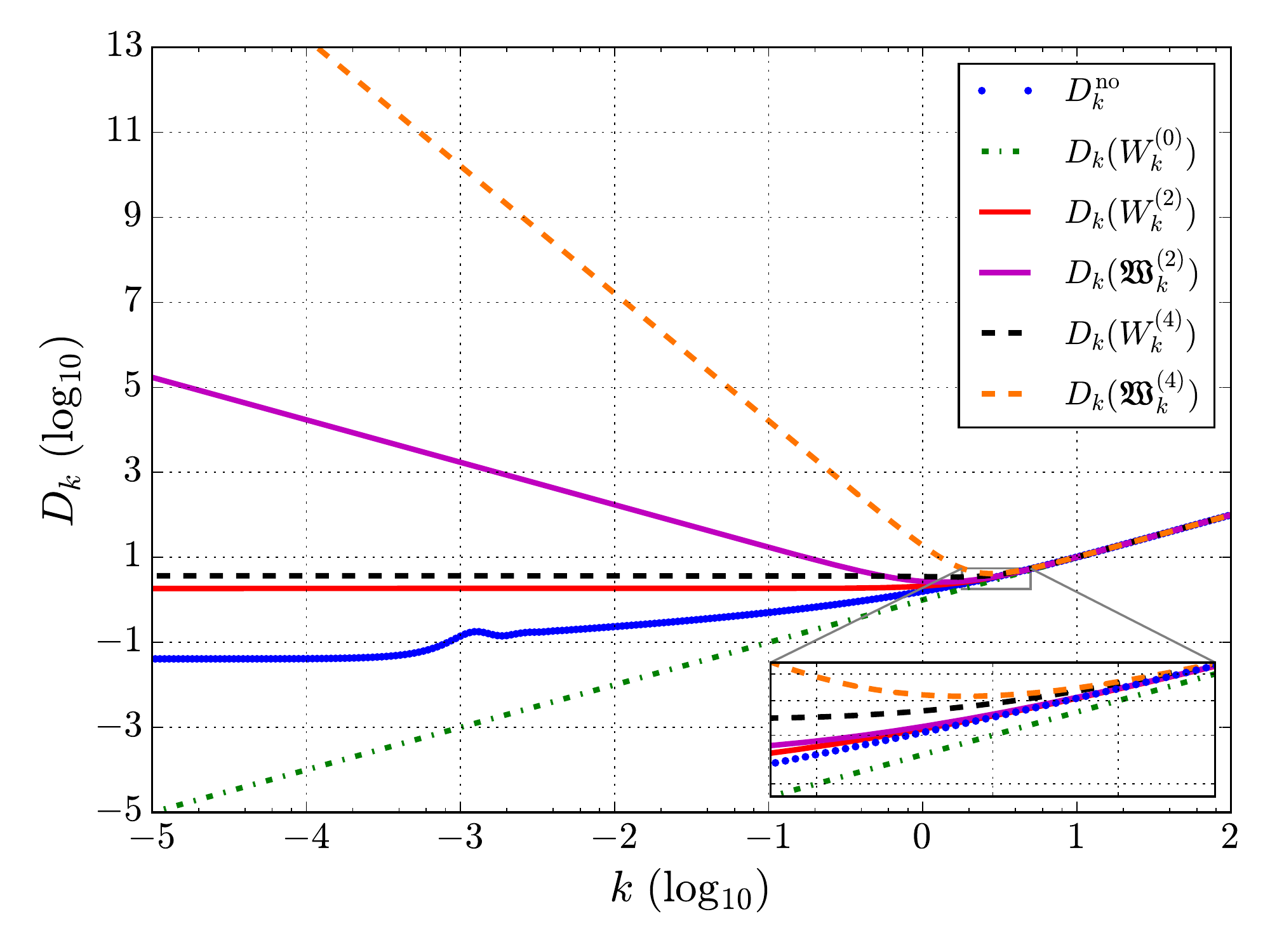} 
\includegraphics[width=0.49\textwidth]{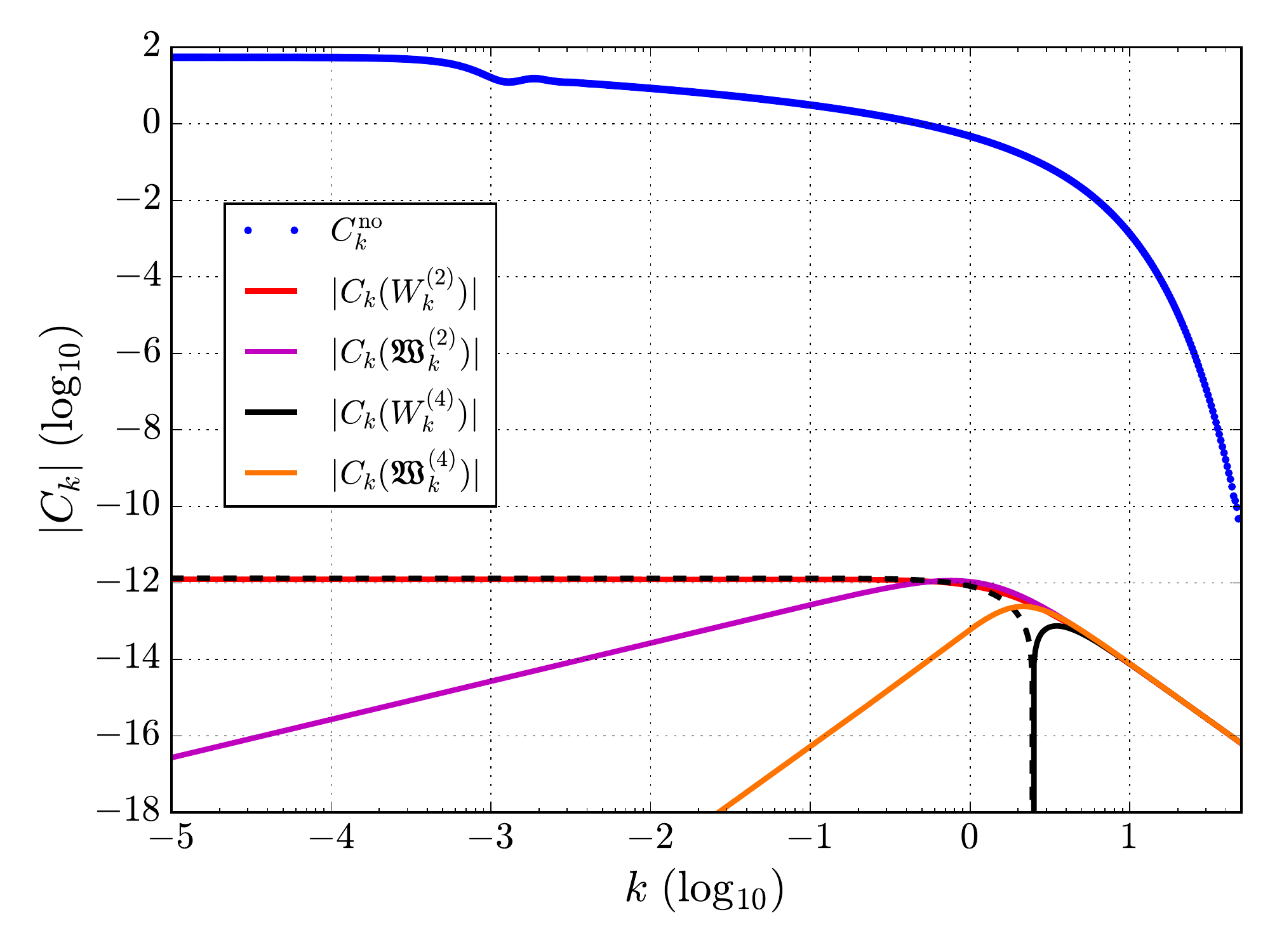}
\caption{Comparison between different vacuum prescriptions: functions that determine the initial conditions of the tensor field, for $\phi_{B} = 0.97$ and $m = 1.20\cdot10^{-6}$. Left panel: function $D_{k}$. Note that $D_{k}(\mathfrak{W}^{(0)}_{k}) = k = D_{k}(W^{(0)}_{k})$. Right panel: function $C_{k}$. Here, dashed lines indicate negative values of $C_{k}$. We note that, for the 0th-order adiabatic initial conditions, $C_k(\mathfrak{W}^{(0)}_{k})= 0 =C_k(W^{(0)}_{k})$.}
\label{fig:Dk-Ck}
\end{figure}
In Fig. \ref{fig:Dk-Ck} we plot these functions for a particular choice of the mass $m$, and of the initial value of the background scalar field, $\phi_B$. Other choices have also been considered, and the corresponding functions $D_k$ and $C_k$ have been checked to show similar behaviors. It is worth noticing that, for all such choices, one always obtains smooth functions of $k$. Besides, all the considered prescriptions lead to functions that agree in the sector of ultraviolet modes, where $D_k\to k$ and $C_k\to 0$. Notice that, for the considered 0th-order adiabatic state, $D_k= k$ and $C_k=0$ exactly for all $k$. In addition, in the interval of wavenumbers $k$ of interest, the function $C_k$ turns out to be negligible for all the analyzed prescriptions except for the non-oscillating vacuum. In this latter case, $C_k$ takes considerably bigger values. Nonetheless, $C_k\to 0$ when $k\to\infty$ in all cases, and for the non-oscillating vacuum this convergence seems to be faster than for the other adiabatic vacua.

The second procedure by which we will compare the different initial conditions is by means of the antilinear coefficients of the Bogoliubov transformations that relate them. Let us start with the set of (orthonormal) complex solutions determined by some given initial data. We will call $\{\mu^{ (r)}_{k}\}_{k\in\mathbb{R}}$ this reference set of solutions, and denote the considered initial data as
\begin{equation}\left\{\left(\mu^{ (r)}_{k,0}, \mu^{\prime{ (r)}}_{k,0}\right)\right\}_{k\in\mathbb{R}}.
\end{equation} Any other (new) set of complex solutions $\{\mu^{(n)}_{k}\}_{k\in\mathbb{R}}$, selected by the initial data 
\begin{equation}\left\{\left(\mu^{ (n)}_{k,0}, \mu^{\prime{ (n)}}_{k,0}\right)\right\}_{k\in\mathbb{R}},
\end{equation} is related to the previous one by a Bogoliubov transformation:
\begin{equation}
\mu^{(n)}_{k} = \alpha_{k}\,\mu_{k}^{(r)} + \beta_{k}\,\mu_{k}^{(r)\ast}, \qquad \qquad |\alpha_{k}|^{2}-|\beta_{k}|^{2} = 1,  \qquad \forall k \in \mathbb{R}.
\end{equation}
The linear and antilinear coefficients are determined by the initial conditions respectively as
\begin{equation}
\alpha_{k} = -i\left[(\mu^{\prime(r)}_{k,0})^{\ast}\mu^{(n)}_{k,0}-\mu^{(r)\ast}_{k,0}\,\mu^{\prime(n)}_{k,0}\right], \qquad \beta_{k} = i\left[\mu^{\prime(r)}_{k,0}\,\mu^{(n)}_{k,0} - \mu^{(r)}_{k,0}\,\mu^{\prime(n)}_{k,0}\right]. 
\end{equation}
Recall that the prime stands for the derivative with respect to the conformal time. The usual physical interpretation of the antilinear coefficients of the Bogoliubov transformation is that the square of their absolute value, $|\beta_{k}|^{2}$, represents the number of particles in a mode $\vec{\bf k}$ (with wavenumber equal to $k$) that contains the vacuum state characterized by the solutions $\{\mu^{(n)}_{k}\}$, as seen in the quantum representation defined by the original vacuum. 

\begin{figure}
\includegraphics[width = 0.5\textwidth]{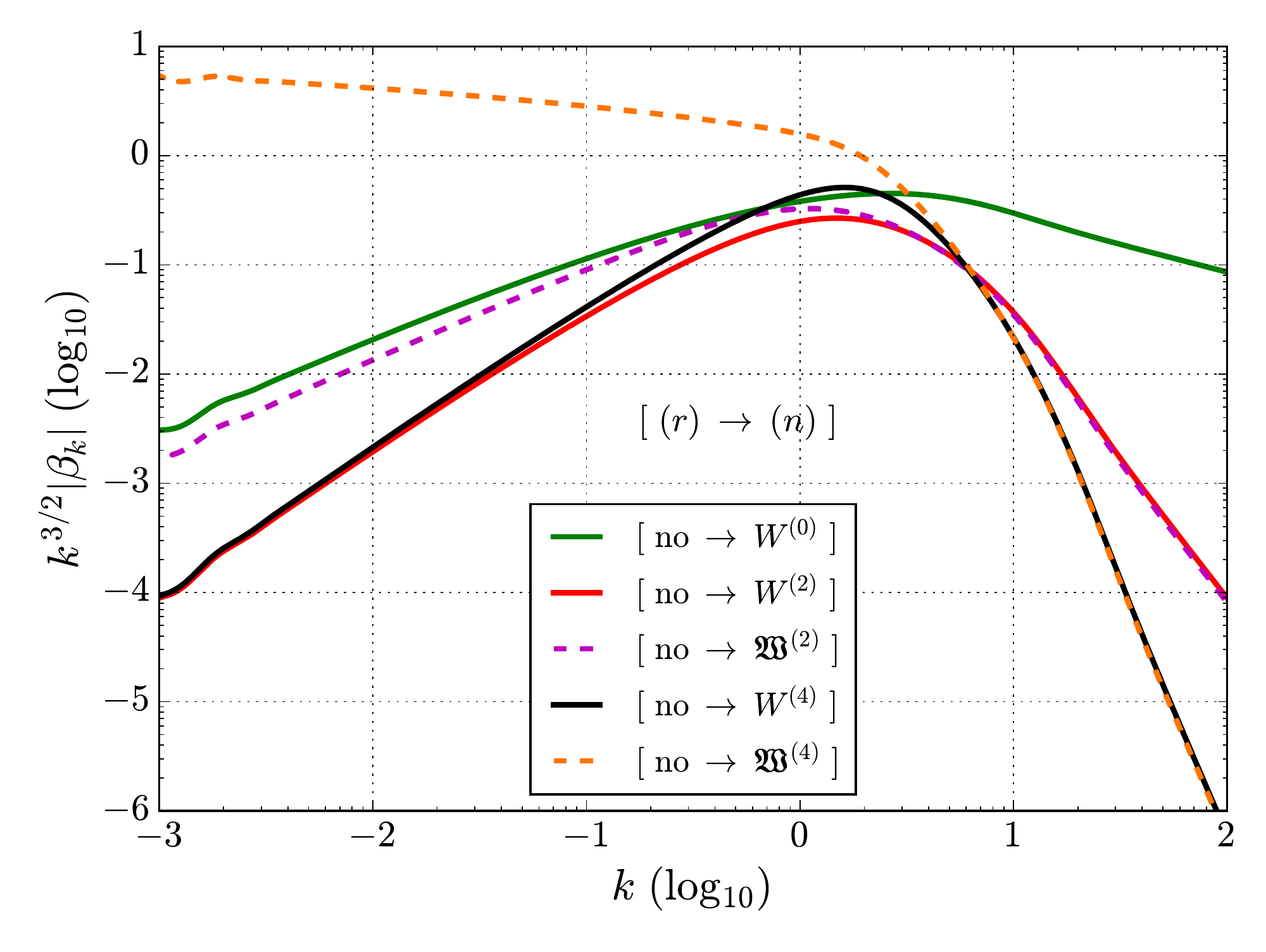}
\includegraphics[width = 0.5\textwidth]{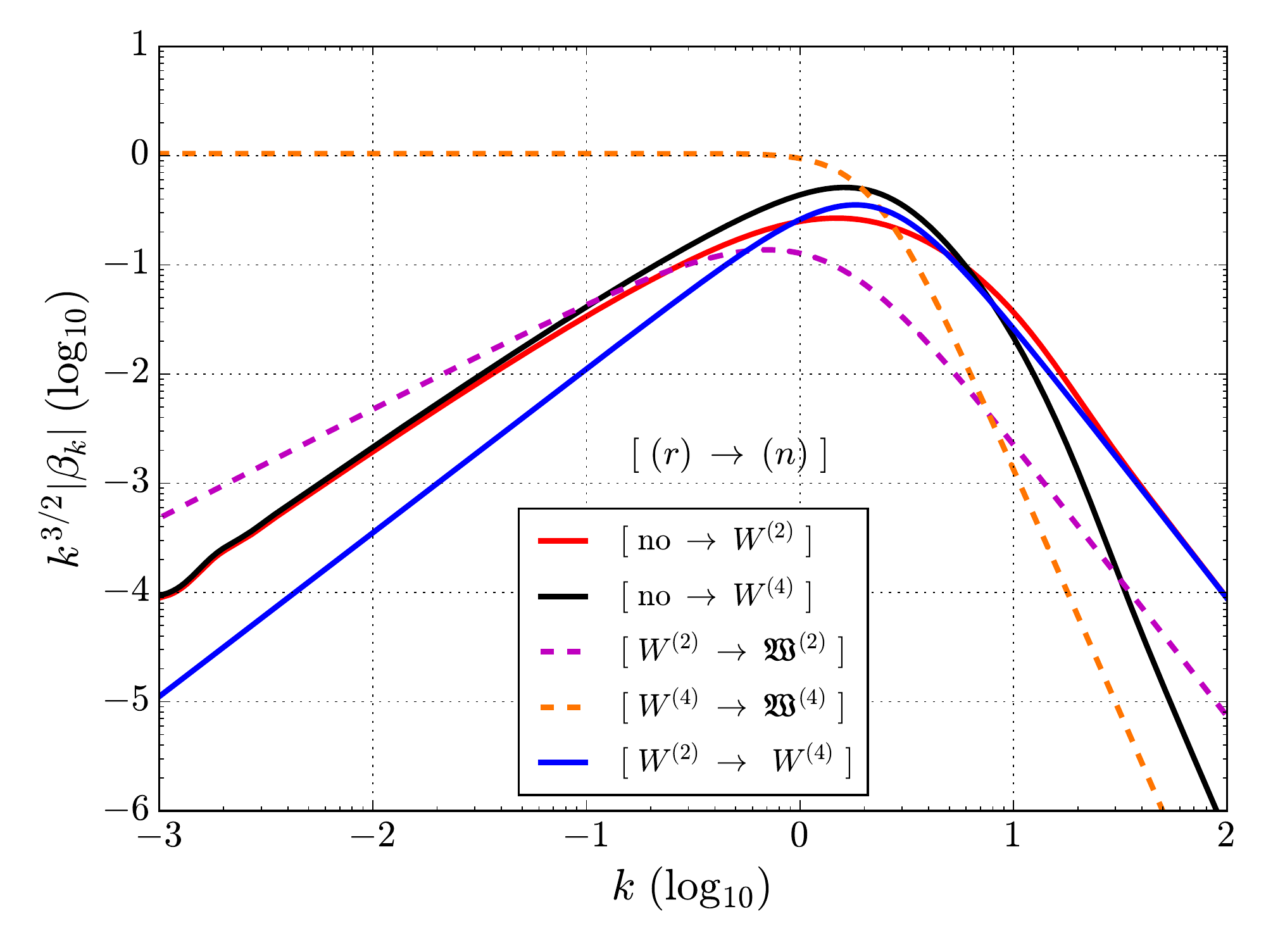}
\caption{Comparison between different vacuum prescriptions for the initial data of the tensor perturbations, for $\phi_{B} = 0.97$ and $m = 1.20 \cdot 10^{-6}$: absolute value of the antilinear coefficients of the Bogoulibov transformation relating the vacuum state of reference, $(r)$, with another vacuum, $(n)$. Left panel: comparison taking the non-oscillating (no) vacuum as the reference state.}
\label{fig:beta}
\end{figure}

As we have commented, for adiabatic initial data of order $\mathfrak{n}$, the functions $W^{(\mathfrak{n})}_{k}$ and $\mathfrak{W}^{(\mathfrak{n})}_{k}$ (as well as any other function of order $\mathfrak{n}$) have the same asymptotic behavior up to terms that are $\mathcal{O}(k^{-1-\mathfrak{n}})$. If we consider an adiabatic state of order $\mathfrak{n}$ as the vacuum of reference, then it is not difficult to realize that the antilinear coefficients $|\beta_{k}|$ corresponding to any other different adiabatic state of order $\tilde{\mathfrak{n}}$ behave for large $k$ as $k^{-2-\mathfrak{m}}$, where $\mathfrak{m}=\min (\mathfrak{n},\tilde{\mathfrak{n}})$ is the minimum of the two adiabatic orders. In other words, the decay of $|\beta_{k}|$ for large $k$ is completely determined by the adiabatic state of lower order. In Fig. \ref{fig:beta} we represent the absolute value of the antilinear Bogoliubov coefficients (multiplied by $k^{3/2}$) obtained with different sets of initial data, providing both the reference vacuum and the final state, with labels $(r)$ and $(n)$, respectively. In the right panel, one can see that the coefficients have the same asymptotic decay for the non-oscillating vacuum state as in the case of the 4th-order adiabatic vacua considered in our discussion, taking one of the latter as reference vacuum. We have also generated adiabatic initial data up to 8th order and compared them numerically with the non-oscillating vacuum. Once more, both sets of initial data happen to lead to Bogoliubov coefficients with the same asymptotic behavior. However, since we are not providing analytic expressions for such adiabatic vacua of 6th and 8th order in this article, we have preferred not to display those numerical results. 

In conclusion, our numerical analysis suggests that the non-oscillating vacuum behaves like a high-order adiabatic state at least up to 8th order, and perhaps up to an even higher order. If this is actually the case, our prescription for the selection of initial data corresponding to the non-oscillating vacuum in fact picks out a state that is tantamount to a high-order adiabatic vacuum, although by means of a procedure that overcomes the potential problems of ill-definiteness that one finds in the usual constructions of adiabatic states.

\section{Comparison with observations}\label{sec:PS-ATPS}

\subsection{Preliminaries}

We will now use the prescriptions that we have proposed above for the choice of vacuum of the perturbations (keeping in mind that other interesting choices do exist) and extract predictions about cosmological observables in the hybrid quantization approach, comparing them with observations. In this section, we will derive the power spectrum of the tensor perturbations, $\mathcal{P}_{\mathcal{T}}$, and compute its spectral index, $n_{t}$, as a function of $k$. Besides, recalling previous results of  Ref. \cite{hybr-pred} for scalar perturbations, we will calculate the tensor-to-scalar ratio, $r = \mathcal{P}_{\mathcal{T}}/\mathcal{P}_{\mathcal{R}} $. Here, $\mathcal{P}_{\mathcal{R}}$ denotes the comoving curvature primordial power spectrum. Finally, we will study possible deviations with respect to the standard results, deduced for perturbations in the Bunch-Davies state at the onset of inflation, paying special attention to the slow-roll consistency relations. 

Different strategies can be followed in order to compute these cosmological observables. One possibility, numerically accurate but computationally expensive, is to evolve the set of tensor modes from the bounce to the end of inflation. This gives the exact value of the power spectrum, up to numerical errors. Alternatively, one can employ the standard (slow-roll) single-field inflation formulation. In this case, although one does not need to evolve the tensor modes, resulting in a procedure which is numerically more efficient, one must take into account that the approximations that are implicit in the adopted formulation reduce the precision at the end of the day. 

Let us describe succinctly the kind of computations that are typical in standard (slow-roll) single-field inflation, before we derive the power spectrum of the tensor perturbations for the different vacua under consideration. In this way, we will be able to compare predictions obtained in different ways, checking the robustness of our results.

It is well known that, for modes that leave the Hubble horizon during inflation in the slow-roll regime, the power spectrum of the scalar and the tensor perturbations can be approximated by the first-order slow-roll expressions  \cite{lang}
\begin{equation}\label{eq:0thSSL}
\mathcal{P}^{(0)}_{\mathcal{R}}(k)=\frac{H^{2}_{\ast}}{\pi \epsilon_{H\ast}},
\end{equation}
and
\begin{equation}\label{eq:0thTSL}
\mathcal{P}^{(0)}_{\mathcal{T}}(k)=\frac{16H^{2}_{\ast}}{\pi},
\end{equation}
where $H=\dot{a}/a$ is the Hubble parameter and $\epsilon_{H} = -\dot{H}/H^{2}$ is the first slow-roll parameter in the Hubble-flow functions. The subscript $\ast$ means that the time-dependent variable is evaluated at $\eta_{\ast}$, defined as the moment when $k = a(\eta_{\ast})H(\eta_{\ast})$, namely, when the studied mode crosses the Hubble horizon. Besides, the dot stands here for the derivative with respect to the proper time. By convention, we will consider that the slow-roll regime starts when the absolute values of the slow-roll parameters $\epsilon_{H}$ and $\eta_{H}=\ddot{H}/(2\dot{H}H)$ are both smaller than $10^{-2}$. The above expressions give relatively good results in such a slow-roll regime. But we must keep in mind that they are only first-order slow-roll formulas. This means that they disregard contributions of higher order in the slow-roll parameters. Those contributions can actually be restored. In fact, together with the power spectra provided by the previous expressions, in this article we will consider more accurate formulas that involve second-order corrections in the slow-roll parameters. In doing so, we follow several strategies. The first one consists in taking the second-order slow-roll expressions and evaluate them at the time when the modes under consideration exit the Hubble horizon (i.e. at the horizon crossing). The second strategy for the computation of the primordial power spectra is based on the previous evaluation, but this time considering only a reference mode, and introducing then a suitable extrapolation from this reference scale to other wavenumbers $k$. We will employ two different extrapolation functionals that are often used in the literature and lead to suitable parameterizations of the primordial power spectra of the scalar and the tensor perturbations \cite{planck-inf,encyc}. 

The first extrapolating expansion we will consider here can be found in Ref. \cite{encyc} (and the references therein). It assumes the following parameterization for the power spectrum of the scalar perturbations:
\begin{equation}\label{eq:PSS-param1}
\frac{\mathcal{P}_{\mathcal{R}}(k;k_{\rm ref})}{\mathcal{P}^{(0)}_{\mathcal{R}}(k_{\rm ref})}=a_0^{\mathcal{R}}(k_{\rm ref})+a_1^{\mathcal{R}}(k_{\rm ref})\ln\left(\frac{k}{k_{\rm ref}}\right)+\frac{a_2^{\mathcal{R}}(k_{\rm ref})}{2}\ln^2\left(\frac{k}{k_{\rm ref}}\right),
\end{equation}
and the same form for the tensor power spectrum, but replacing the ratio on the left-hand side with $\mathcal{P}_{\mathcal{T}}(k;k_{\rm ref})/\mathcal{P}^{(0)}_{\mathcal{T}}(k_{\rm ref})$, and the coefficients $a_i^{\mathcal{R}}$ with other coefficients $a_i^{\mathcal{T}}$ ($i=0,$ 1, and 2). Both collections of coefficients are functions of time, and the notation $a_i^{\mathcal{R}}(k_{\rm ref})$ and $a_i^{\mathcal{T}}(k_{\rm ref})$ indicates that they must be evaluated at the instant when the mode $k_{\rm ref}$ exits the horizon. In the approximation that we will consider here, these coefficients are  truncated at second order in the slow-roll parameters expansion. For explicit expressions of all these coefficients in terms of Hubble-flow functions, see Eqs. (2.18)-(2.25) of Ref. \cite{encyc}. The wavenumber $k_{\rm ref}$ is an arbitrary reference scale that must exit the Hubble horizon during the slow-roll regime. Note that each of the expressions $\mathcal{P}_{\mathcal{R}}(k;k_{\rm ref})$ and $\mathcal{P}_{\mathcal{T}}(k;k_{\rm ref})$, for the power spectrum of the scalar and the tensor perturbations respectively, is a parabola in the logarithm of $k$. We expect that different choices of $k_{\rm ref}$ will yield slightly different power spectra. On the other hand, if one is not extrapolating, but instead tracking the exit of all the modes of interest, the second-order slow-roll formulas reduce simply to
\begin{equation}\label{eq:2thSSL}
\mathcal{P}^{(2)}_{\mathcal{R}}(k)=\left(a_0^{\mathcal{R}}\right)_\ast\mathcal{P}^{(0)}_{\mathcal{R}}(k)
\end{equation}
and
\begin{equation}\label{eq:2thTSL}
\mathcal{P}^{(2)}_{\mathcal{T}}(k)=\left(a_0^{\mathcal{T}}\right)_\ast\mathcal{P}^{(0)}_{\mathcal{T}}(k).
\end{equation}
Again, the asterisk stands for the evaluation of the time-dependent coefficients $a_0^{\mathcal{R}}$ and $a_0^{\mathcal{T}}$ (truncated at second order) when the mode $k$ crosses the horizon during slow roll.

The second and last strategy that we will consider in this article follows the very same idea, but adopts a different expansion of the power spectra, namely, the one used by the Planck Collaboration in Ref. \cite{planck-inf}. In this case the expansion is given by 
\begin{equation}\label{eq:PSS-param2}
\mathcal{P}^{\rm Pl}_{\mathcal{R}}(k;k_{\rm ref})=A_s(k_{\rm ref})\,\left(\frac{k}{k_{\rm ref}}\right)^{n_s(k_{\rm ref})-1+\frac{1}{2}\ln\left(\frac{k}{k_{\rm ref}}\right)\frac{dn_s}{d\,\ln\, k}(k_{\rm ref})},
\end{equation}
and
\begin{equation}\label{eq:PST-param2}
\mathcal{P}^{\rm Pl}_{\mathcal{T}}(k;k_{\rm ref})=A_t(k_{\rm ref})\,\left(\frac{k}{k_{\rm ref}}\right)^{n_t(k_{\rm ref})+\frac{1}{2}\ln\left(\frac{k}{k_{\rm ref}}\right)\frac{dn_t}{d\,\ln\, k}(k_{\rm ref})}.
\end{equation}
These expressions can be found in Eqs. (5) and (6) of Ref. \cite{planck-inf}. The spectral indices $n_s$ and $n_t$, and their derivatives (runnings) $dn_s/d\,\ln\, k$ and $dn_t/d\,\ln\, k$, are polynomials of second order in the Hubble-flow functions. As it  happens with the coefficients $A_s$ and $A_t$, they depend on the reference mode $k_{\rm ref}$. Their explicit expressions are given in Eqs. (8)-(11) of Ref. \cite{planck-inf}. 

\begin{figure}
\centering
\includegraphics[width=0.49\textwidth]{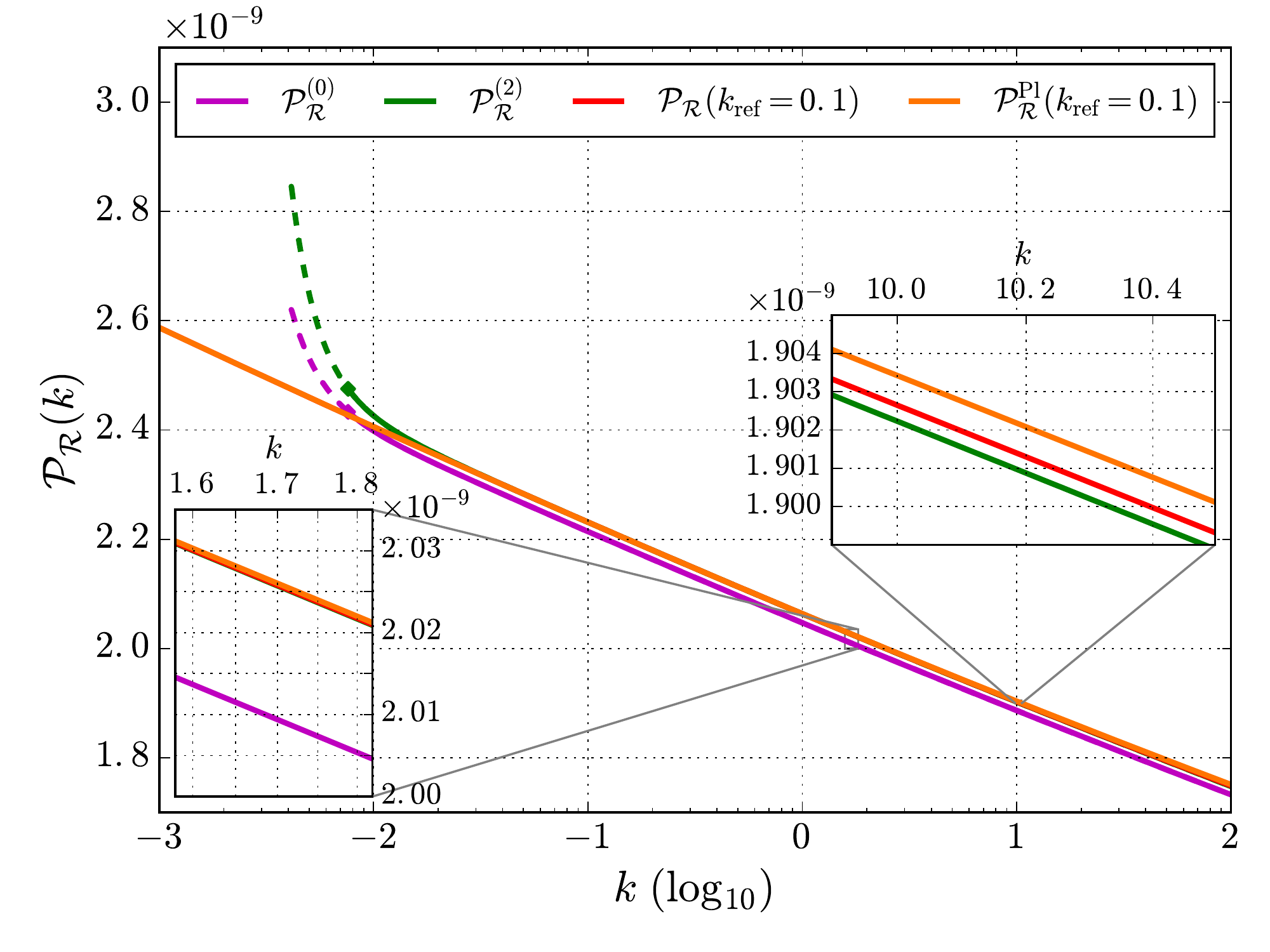} 
\includegraphics[width=0.49\textwidth]{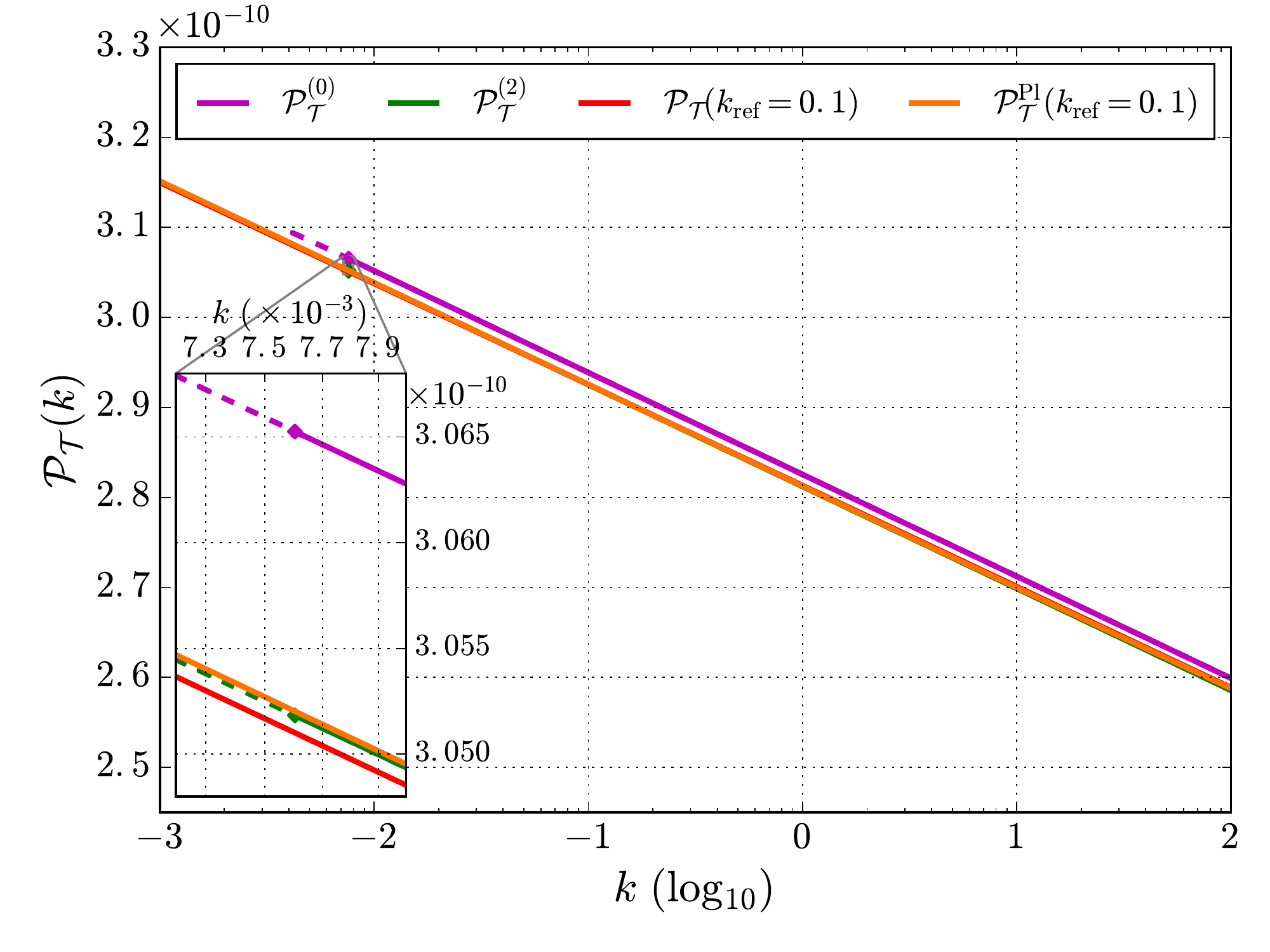}
\caption{Comparison between power spectra calculated with different strategies in the slow-roll regime. Here, $\phi_{B} = 0.97$ and $m = 1.20\cdot10^{-6}$. A dashed line indicates modes that exit the Hubble horizon in an inflationary phase but not in slow-roll regime. Left panel: slow-roll power spectra for the scalar perturbations. Right panel: slow-roll power spectra for the tensor perturbations.}
\label{fig:stndr-SL}
\end{figure}

In Fig. \ref{fig:stndr-SL} we show a generic example of the power spectra of the scalar and the tensor perturbations obtained using the different strategies that we have mentioned above. As we see, all computations are in good agreement for scales that exit the Hubble horizon well into the slow-roll regime, although the first-order formulas \eqref{eq:0thSSL} and \eqref{eq:0thTSL} are not as accurate as the second-order ones, with relative discrepancies of about $0.5\%$. For instance, the second-order formulas give slightly more power in the scalar spectrum and slightly less power in the tensor case. Therefore, the value of the tensor-to-scalar ratio obtained with the first-order formulas is slightly bigger than the actual value, increase which, in this particular model, goes against the observational bounds on this parameter. Moreover, this approximation is the source of the small discrepancies that were already discussed in Ref. \cite{hybr-pred} for the scalar perturbations. With this panorama, we can be confident that our numerical calculations are robust and provide accurate estimations of the power spectrum of the perturbations.

\subsection{Primordial power spectrum of the tensor modes}

We will now focus our attention on the primordial power spectrum of the tensor modes that follows from the hybrid quantization approach adopted in our discussion. For these tensor modes, we will compare the numerical results, obtained evolving the perturbations from the bounce to the end of inflation, with the power spectra computed by using the second-order slow-roll formulas in which one either tracks the exit of all the relevant modes [given in expressions \eqref{eq:2thSSL} and \eqref{eq:2thTSL}], or employs the expansion around a reference scale [given in Eqs. \eqref{eq:PSS-param1}, \eqref{eq:PSS-param2}, and \eqref{eq:PST-param2}]. In addition, we will consider all the different prescriptions that we detailed above for the choice of initial state of the tensor perturbations at the bounce. The same computations were carried out for the scalar perturbations in Ref. \cite{hybr-pred} and will not be repeated here. 

\begin{figure}
\centering
\includegraphics[width=0.49\textwidth]{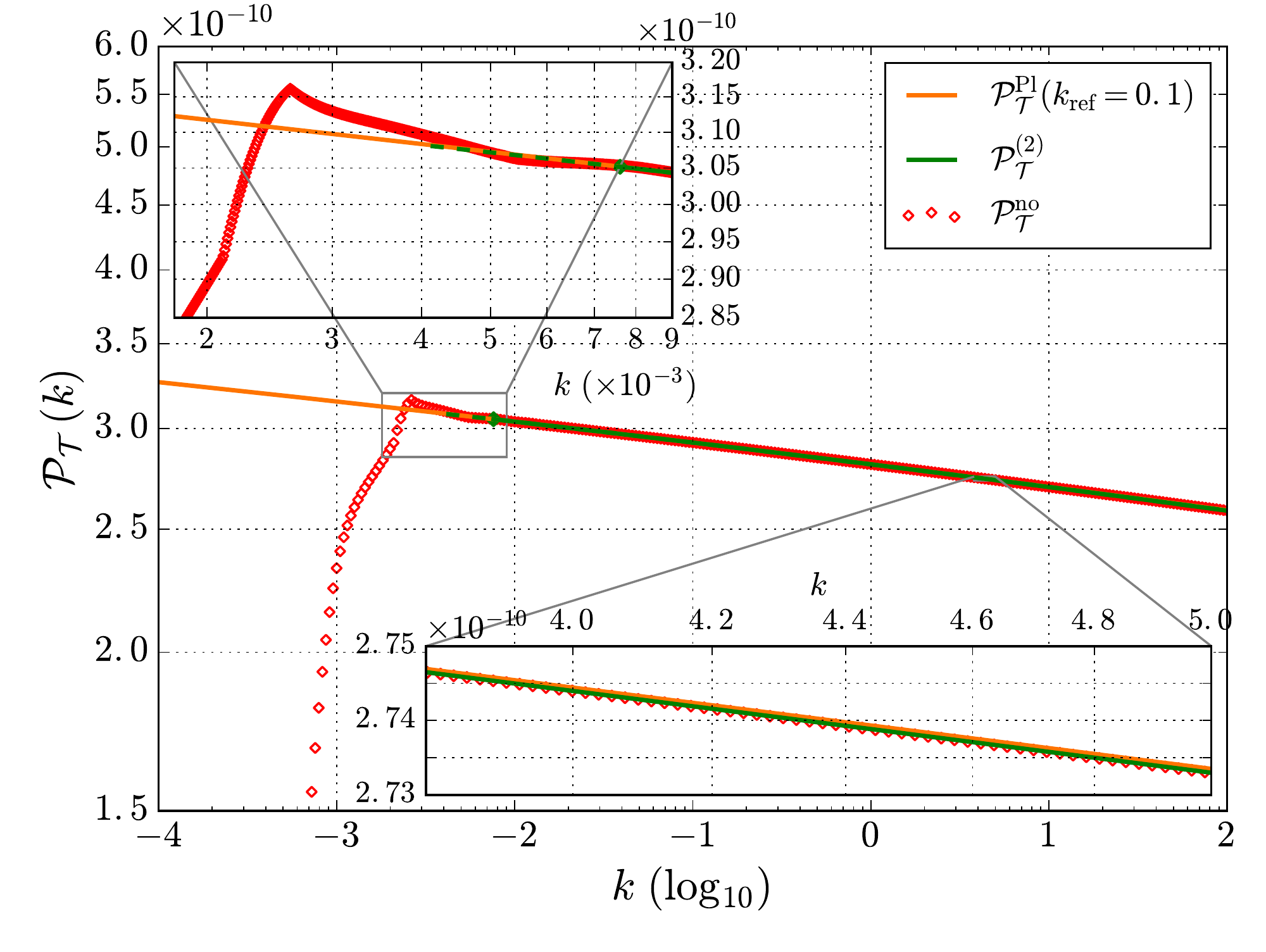} \includegraphics[width=0.49\textwidth]{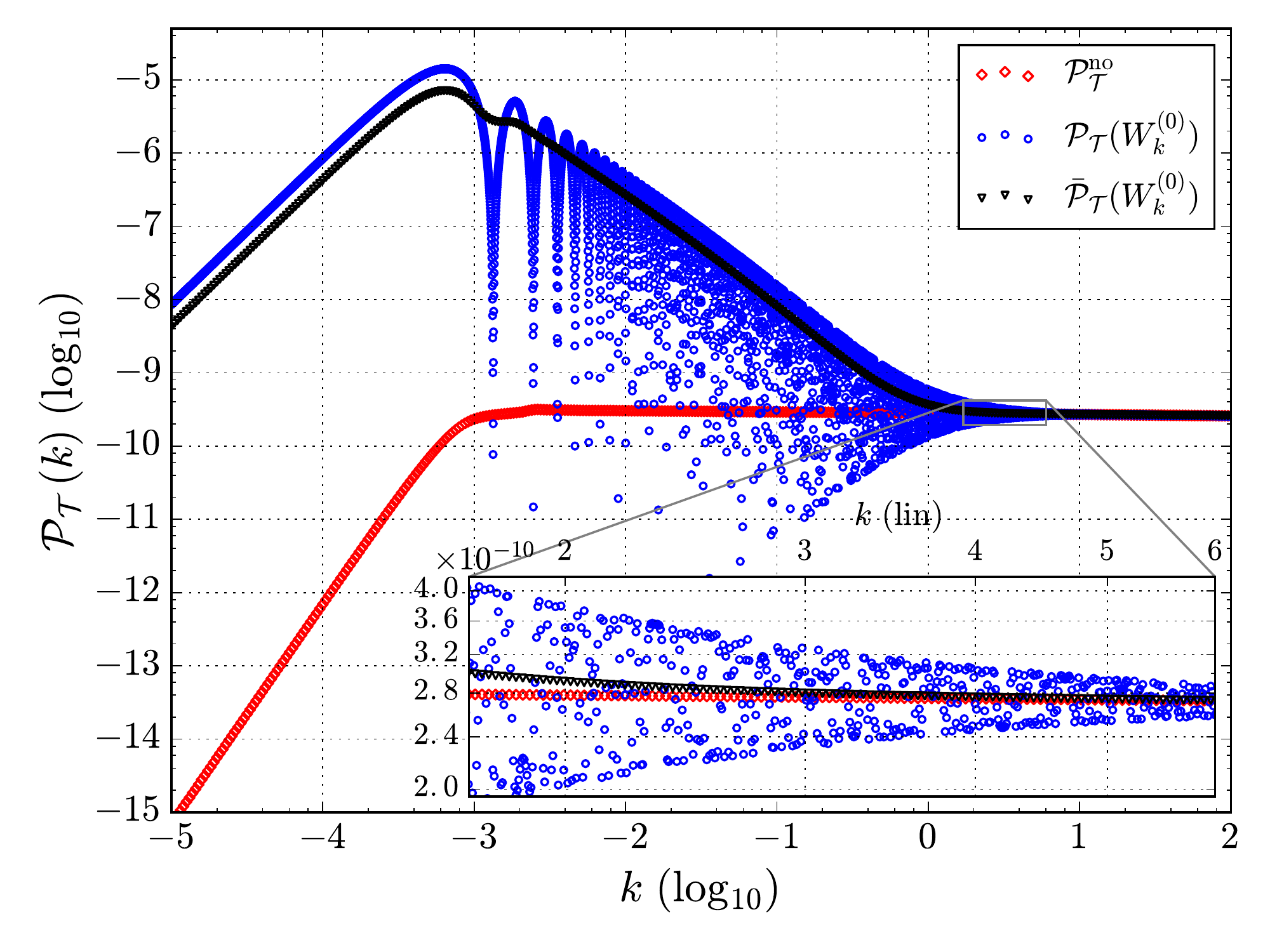}\\
\includegraphics[width=0.49\textwidth]{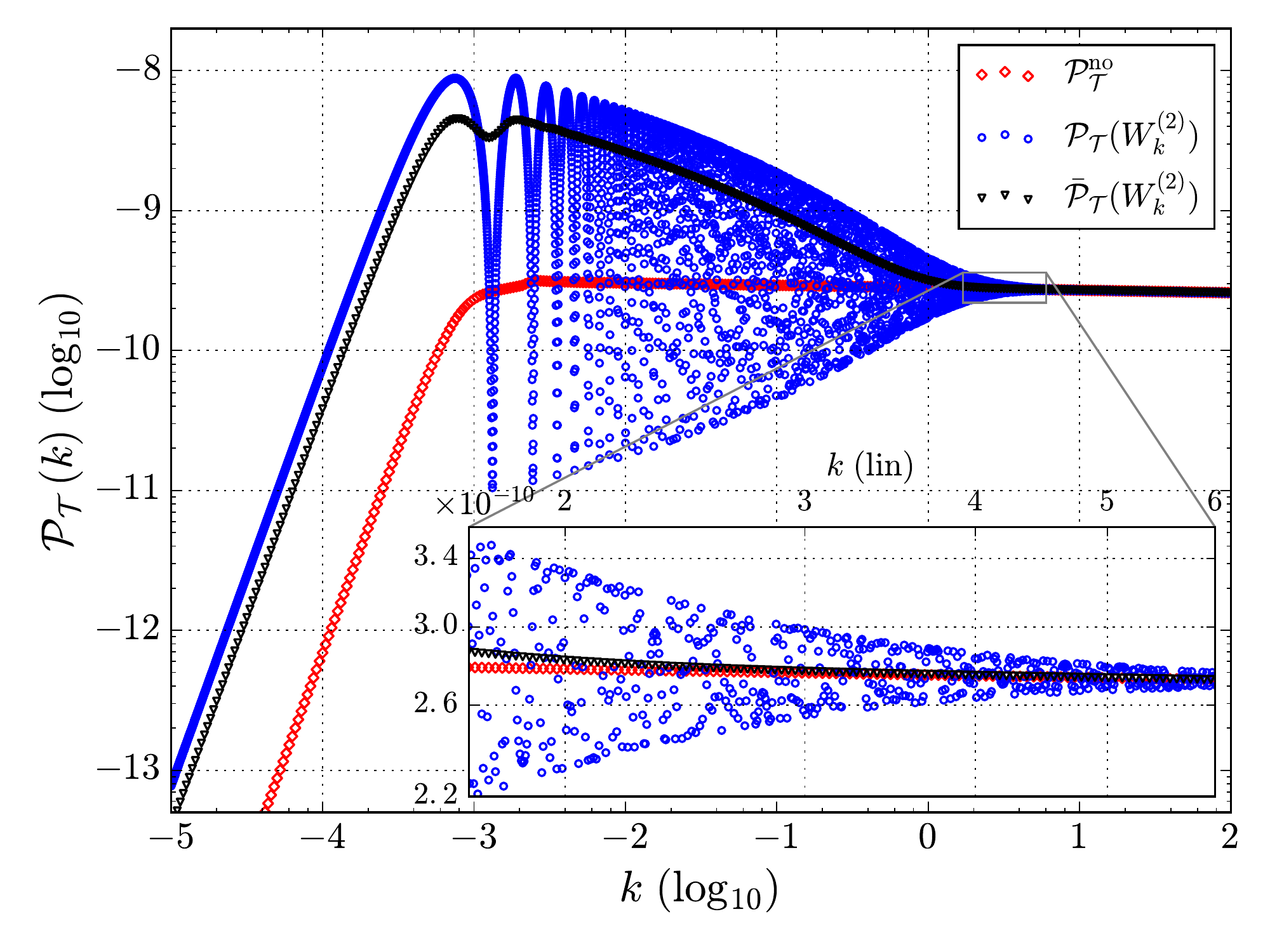} \includegraphics[width=0.49\textwidth]{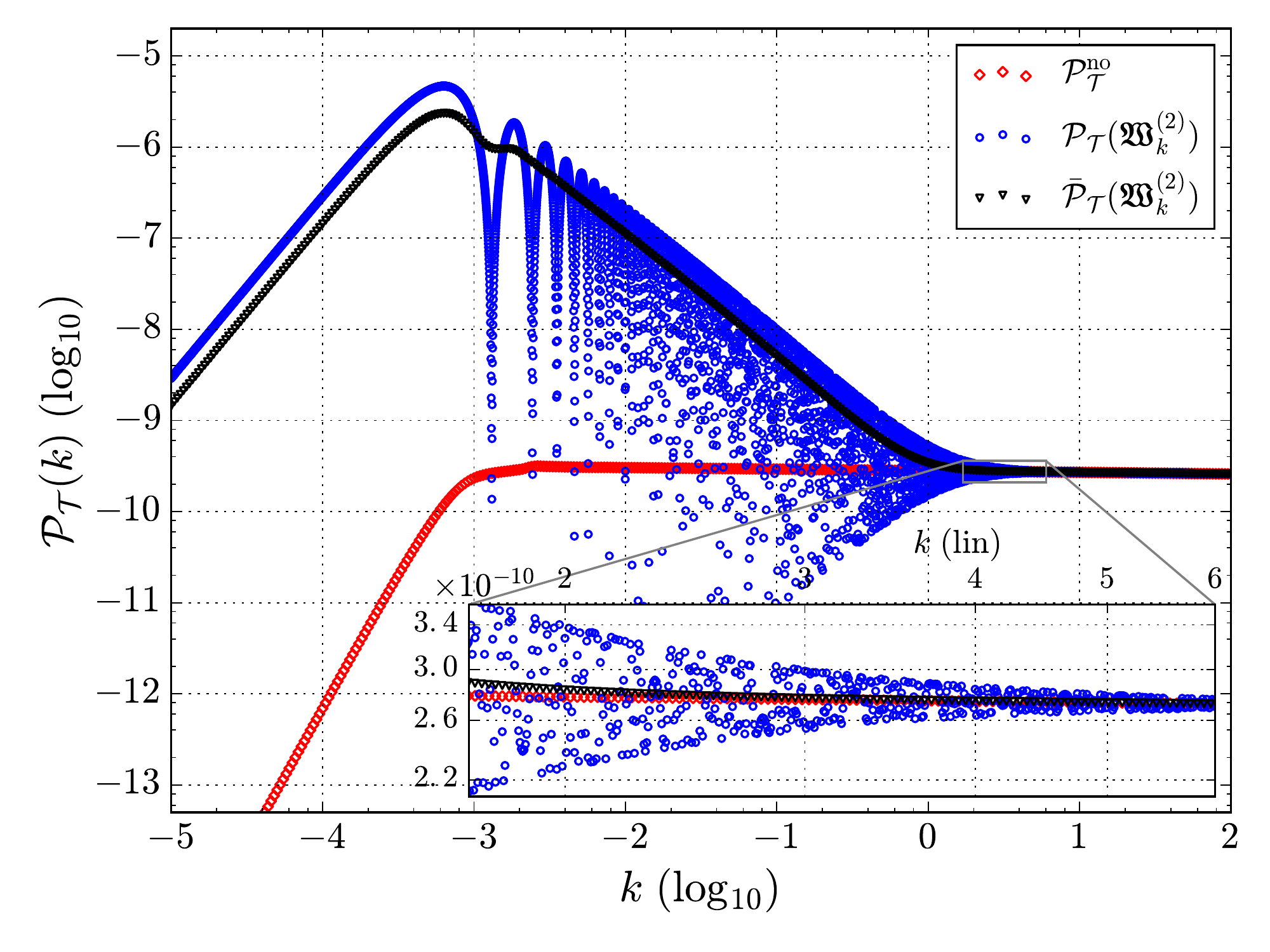}\\
\includegraphics[width=0.49\textwidth]{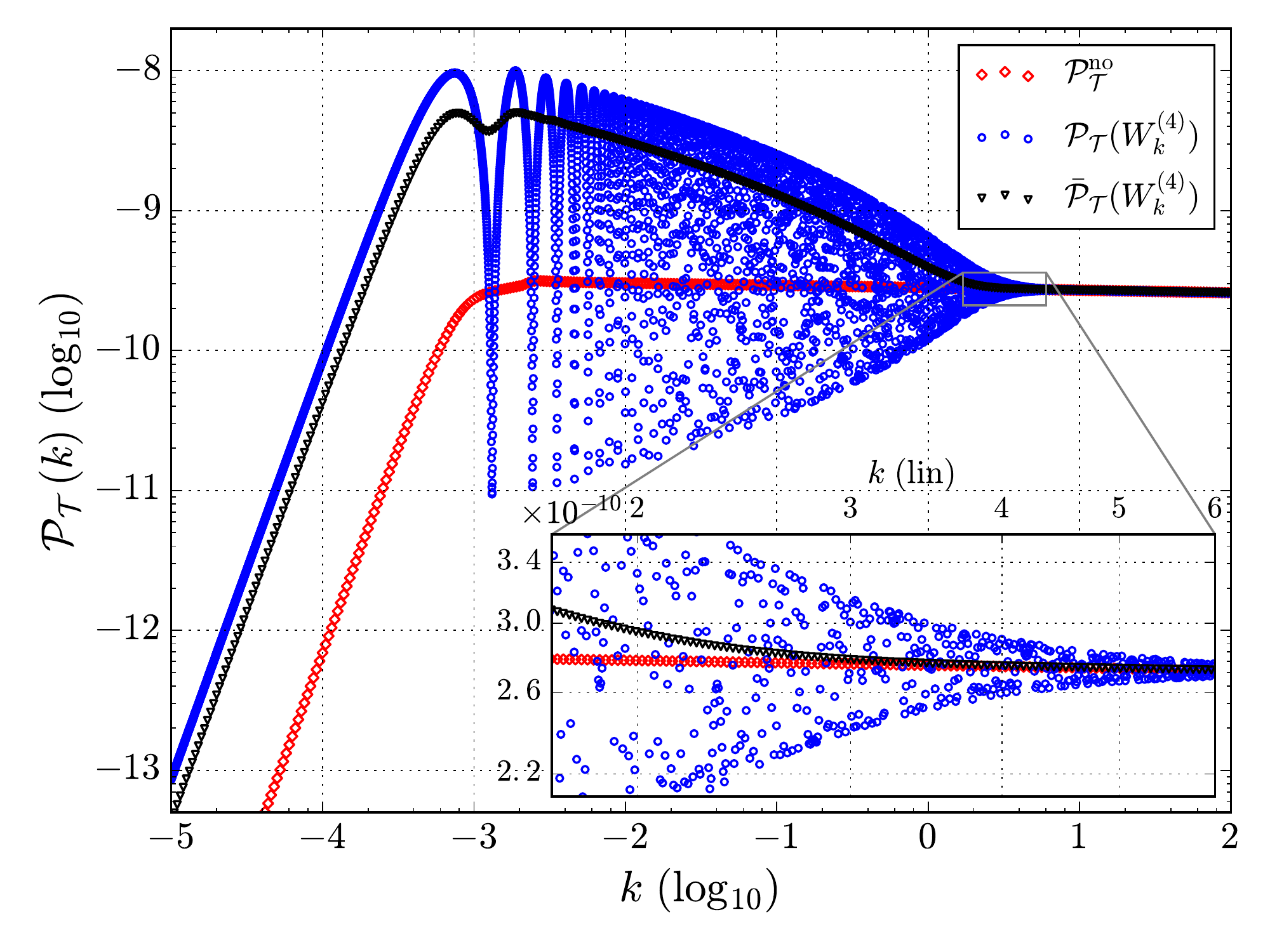} \includegraphics[width=0.49\textwidth]{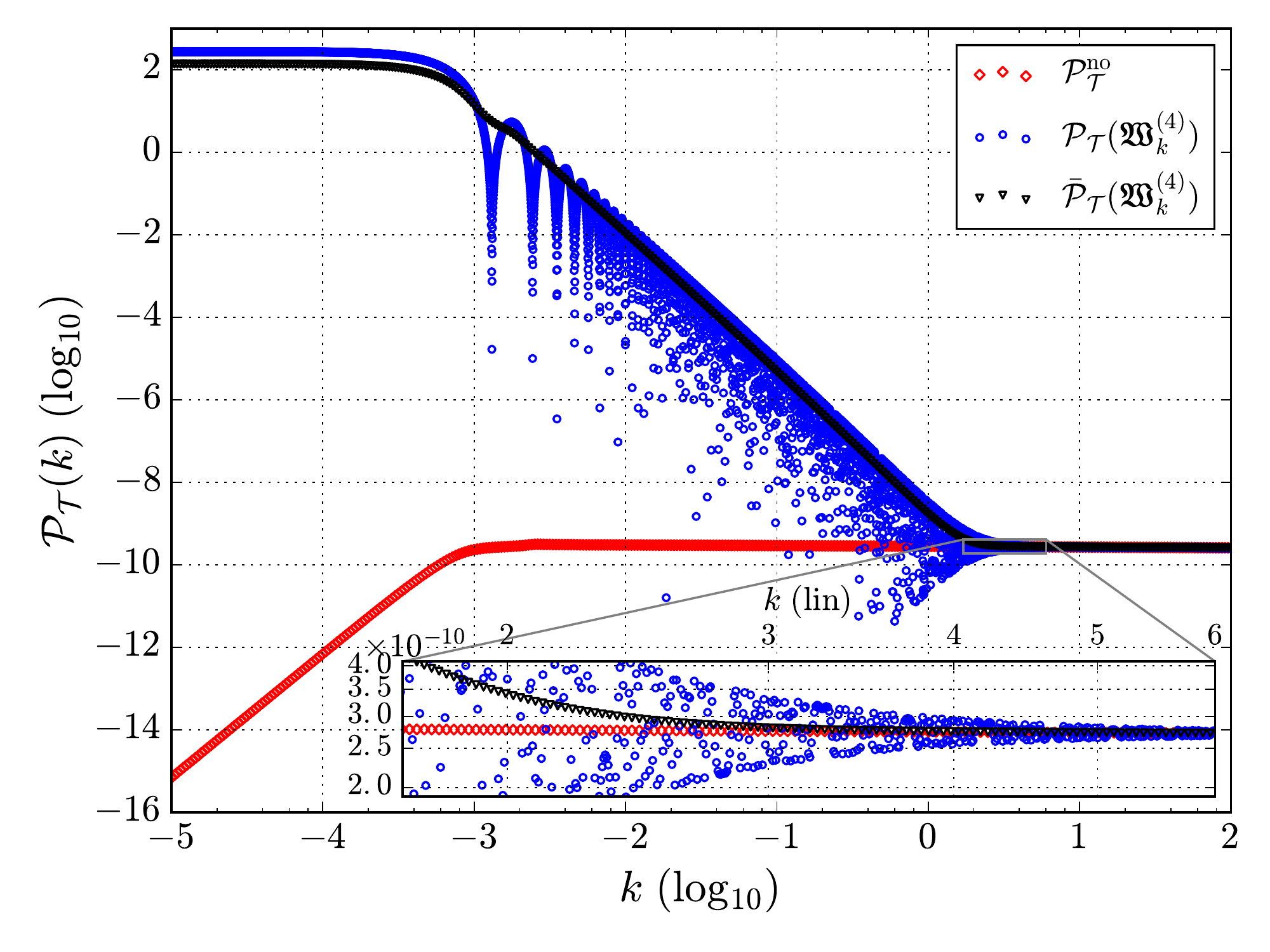}
\caption{Comparison between primordial power spectra obtained with different sets of solutions for the perturbations. Here, $\phi_{B} = 0.97$ and $m = 1.20\cdot10^{-6}$.}
\label{fig:PPTs}
\end{figure}
In the left upper panel of Fig. \ref{fig:PPTs}, we plot the power spectra obtained either with non-oscillating initial conditions, denoted by $\mathcal{P}^{\rm no}_{\mathcal{T}}$, with the slow-roll formula \eqref{eq:2thTSL}, or with the extrapolation functional \eqref{eq:PST-param2}. They all agree very well for modes that cross the Hubble horizon within the slow-roll regime. Nevertheless, those modes that exit the Hubble horizon before the slow-roll phase, have a power spectrum with significantly less power in the case of the non-oscillating vacuum than when it is estimated using the extrapolation \eqref{eq:PST-param2}. In fact, this power spectrum displays two regions with different behaviors (with respect to the slow-roll formula), as it is known that happens as well for the scalar perturbations \cite{hybr-pred}: (i) small oscillations for $  2{.}5\cdot 10^{-3} \lessapprox k \lessapprox 10^{-2}$, with a moderate enhancement and (ii) power suppression for $k \lessapprox 2{.}5\cdot 10^{-3}$, that becomes stronger as $k$ decreases. This strong suppression for large scales is very similar (maybe not surprisingly) to the one found in scenarios where the slow-roll regime is preceded by a kinetically dominated era \cite{pow-supp} (as it is also the case studied here). In any case, we expect the presence of genuine modifications in the spectrum, originated in the earlier phase in which the LQC corrections dominate. 

The primordial power spectra resulting from adiabatic initial conditions are displayed in the rest of panels of Fig. \ref{fig:PPTs}. They have been computed starting with the power spectrum of the non-oscillating vacuum, by taking into account the Bogoliubov transformation that relates this vacuum with the adiabatic ones. The corresponding primordial power spectrum is given by 
\begin{equation}
\mathcal{P}_{\mathcal{T}} (k)= \left[1 + 2|\beta_k|^2 + 2|\alpha_k||\beta_k|\cos\left(\varphi^\alpha_{k}-\varphi^\beta_{k} + 2\varphi^{({\rm no})}_{k}\right)\right] \mathcal{P}^{\rm no}_{\mathcal{T}}(k), 
\end{equation}
where we have used $\alpha_k = |\alpha_k|e^{i\varphi^\alpha_{k}}$, $\beta_k = |\beta_k|e^{i\varphi^\beta_{k}}$, and $\mu^{({\rm no})}_k = |\mu^{({\rm no})}_k|e^{i\varphi^{({\rm no})}_{k}}$. Since the strong oscillations are produced by the term containing the cosine, we have also plotted the primordial power spectrum obtained by setting this term equal to zero. This power spectrum, that we have called $\bar{\mathcal{P}}_{\mathcal{T}}$, gives a good approximation to the result of averaging the highly oscillatory spectrum in small bins in $k$, and also in small bins in $\ln k$ for $k \gtrapprox 10^{-2}$. For the primordial power spectra of adiabatic states, we can distinguish three regions with different behavior. In the first one, formed by large wavenumbers $k$, the behavior is similar to that of the slow-roll formula, and consequently similar to the non-oscillating spectrum. The second region covers the interval $10^{-3}\lessapprox k\lessapprox 4$. Here we observe high oscillations. This region is usually interpreted as governed by particle production processes, owing to the fact that the corresponding modes exit and reenter the Hubble horizon in the pre-inflationary regime. Finally, in the third region, which runs over $k\lessapprox 10^{-3}$, one gets a suppression of power for $W^{(0)}_{k}$, $W^{(2)}_{k}$, $\mathfrak{W}^{(2)}_{k}$, and $W^{(4)}_{k}$, but a large and approximately constant power for $\mathfrak{W}_{k}^{(4)}$. For all of them, the existence of big oscillations produce in average a large enhancement in the power spectrum that is not compatible with present observations, unless the involved scales are not currently observable in the CMB. We have checked numerically that the presence of these big oscillations and the associated enhancement of power for the adiabatic states are robust results, independent of the choice of the bounce as the initial time surface. Namely, the same kind of qualitative results are obtained if one changes the instant where the initial conditions that determine the adiabatic vacuum are imposed, moving this initial instant away from the bounce, as far as it is not chosen very close to the onset of inflation. For these reasons, and owing to the fact that adiabatic vacuum states have already been studied within LQC in several references (see for instance \cite{AAN1,AAN2,agu-morr}), in this work we will mostly concentrate our attention on the non-oscillating vacuum. 

The Planck Collaboration has not been able to detect primordial tensor perturbations, e.g. by carrying out accurate measurements of the $B$-modes polarization. However, it has been possible to provide bounds on the tensor-to-scalar ratio 
\begin{equation}
r=\frac{\mathcal{P}_{\mathcal{T}}}{\mathcal{P}_{\mathcal{R}}}.
\end{equation}
Here, we have obviated the $k$-dependence of the different involved quantities. The mentioned bounds, which correspond to a comoving scale of $0{.}002{\rm Mpc}^{-1}$, are given by $r_{0{.}002}<0{.}10$ (95\%CL, Planck TT+lowP) and $r_{0{.}002}<0{.}11$ (95\%CL, Planck TT+lowP+lensing)\footnote{CL stands for confidence level. Each bound is based on a different set of observations which are described in the legend inside the parentheses, and that are explained in Ref. \cite{planck-inf}.}. To derive these bounds, the Planck Collaboration mostly assumes the validity of the first-order slow-roll consistency relation $r = -8n_t$, although they also use the second-order relation $n_t = -r(2 - r/8 - n_s)/8$ when they consider the possibility of a running. 

\begin{figure}
\centering
\includegraphics[width=\textwidth]{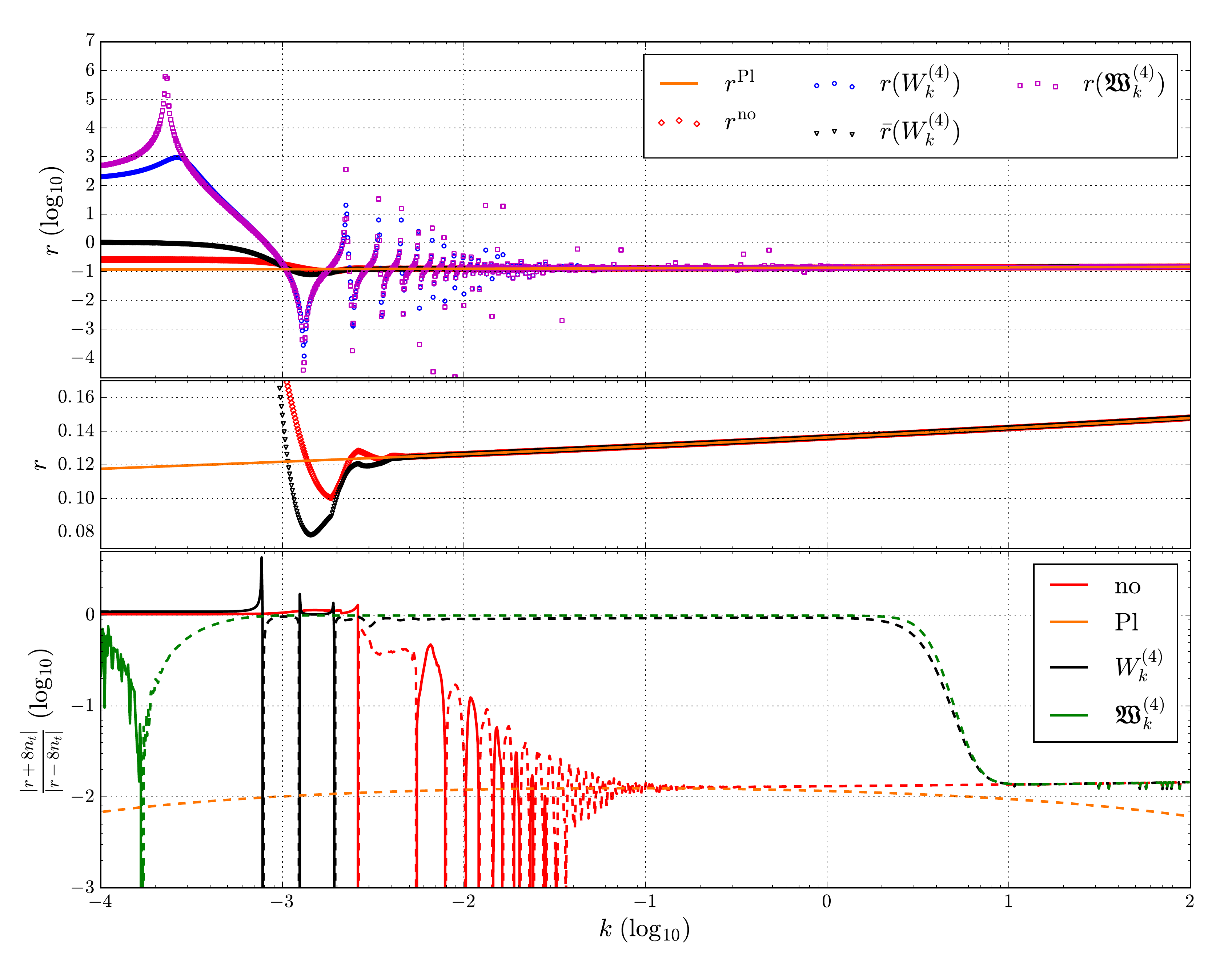} 
\caption{Comparison between different vacuum prescriptions: tensor-to-scalar ratio and validity of the consistency relation $r = - 8 n_t$. In the two upper panels, we show the tensor-to-scalar ratio for the non-oscillating vacuum and for the two 4th-order adiabatic vacua considered in the text. The ratio for these adiabatic vacua has been computed both from the full power spectrum and from its averaged version. Since it turns out that $\bar{r}(W^{(4)}_{k}) \approx \bar{r}(\mathfrak{W}^{(4)}_{k})$, we display explicitly only the former of these ratios. In the lowest panel, we plot the quantity $|r + 8n_t|/|r - 8n_t|$. Dashed (solid) lines indicate negative (positive) values of $r + 8n_t$. For both $W^{(4)}_{k}$ and $\mathfrak{W}^{(4)}_{k}$, we plot the results using the averaged power spectra. Here, we have taken $\phi_{B} = 0.97$ and $m = 1.20\cdot10^{-6}$.}
\label{fig:r-vs-nt}
\end{figure}

For primordial power spectra computed numerically, we have looked at the spectral index for the tensor perturbations locally, defined as $n_t = d \ln \mathcal{P}_{\mathcal{T}}/d \ln k$. Nonetheless, for scales in which the power spectrum oscillates rapidly, it seems much more natural to use the averaged spectrum in order to compute the tensor-to-scalar ratio and the spectral index. This ratio will be called $\bar{r}$. In Fig. \ref{fig:r-vs-nt} we show the tensor-to-scalar ratios for the non-oscillating vacuum and the two 4th-order adiabatic vacua considered in our discussion, along with the relative difference between $r$ and $-8n_t$, that immediate tells us the sector of wavenumbers $k$ where the consistency relation is violated. Several comments are in order. First, for the inflationary model and the parameters considered here, the consistency relation is not satisfied exactly even for scales that exit the Hubble horizon during the slow-roll regime. For those scales, the value of $-8n_t$, with $n_t$ obtained by numerical differentiation, turns out to be about $2\%$ higher than the actual value of the tensor-to-scalar ratio. This relative discrepancy can be reduced to $1\%$ for large wavenumbers $k$ when the power spectra are computed using the expressions \eqref{eq:PSS-param2} and \eqref{eq:PST-param2}. Second, the consistency relation is violated and does not even give approximate results at scales around or smaller than the one where the strong oscillations start in the power spectrum. For these scales with oscillations, as we have already discussed, there is in average an increase in power, both for the scalar and the tensor perturbations. From Fig. \ref{fig:r-vs-nt}, it is clear that this enhancement is the same for both kinds of perturbations, given that the tensor-to-scalar ratio is approximately the same as the one calculated with the second-order slow-roll formulas. Nonetheless, such enhancement modifies significantly the value of the spectral index, making it more negative. A final comment concerns the non-oscillating vacuum. In this case, the consistency relation gives a fairly good approximation to the tensor-to-scalar ratio for scales exiting the horizon during slow roll; however, the relation does not hold for scales that show a strong suppression. In fact, for these latter scales, the tensor-to-scalar ratio depends significantly on the phases at which the tensor perturbations and the scalar ones get frozen at the horizon crossing. For the parameters explored in this particular inflationary model, we always obtain a larger tensor-to-scalar ratio in this region of large scales.

\subsection{CMB polarization: $TT$, $TE$, $EE$, and $BB$ correlation functions}

We will now compare the predictions obtained with our hybrid quantization approach in the case of the non-oscillating vacuum with observations of the Planck Collaboration. Actually, in order to do this, we need first to perform a \emph{scale matching}. In our numerical simulations, we have arbitrarily fixed the volume at the bounce as $v_{B} = 1$. Nevertheless, for the observations registered by the Planck Collaboration, the convention consists in fixing the scale factor today by setting $v_{o} = 1$ (the subindex $o$ denotes evaluation at present), as it is usually done in the cosmology literature. Therefore, one must provide a correspondence between the comoving scales $k$ and the physical scales of the observations. With this aim, we will follow this procedure: we will take the value of the power amplitude $A_{s}$ observed by Planck at the pivot mode $k_{\ast} = 0.05\, {\rm Mpc}^{-1}$, find the corresponding scale $k_{\star}$ at which our theoretical value of the primordial power spectrum coincides with $A_{s}$, and identify this latter scale with $k_{\ast}$. Namely, we will adjust the scale so that $\mathcal{P}^{\rm no}_{\mathcal{R}}(k_{\star})=A_{s}$. It is worth recalling that the amplitude of the power spectrum given by the Planck Collaboration is inferred from the observational data by adopting a parameterization of the form \eqref{eq:PSS-param2}, motivated by the slow-roll approximation. Since the primordial power spectra studied here are not monotonous functions of $k$ in the intervals under consideration, it might happen that there exists more than one scale that satisfies our matching condition. Nonetheless, the power spectra are in fact monotonous in the region of scales that are well inside the slow-roll regime at the horizon crossing. So, we will consider only those scales to perform the scale matching, choosing in this way $k_{\star}$ as a mode that exits the horizon definitively in the slow-roll phase. The resulting value of the pivot scale $k_{\star}$ will depend both on the mass $m$ of the scalar field and on the value $\phi_{B}$ of its homogeneous mode at the bounce. 

We will compare our results with the best-fit curve provided by the Planck Collaboration \cite{planck} (which is given for the TT+lowP data). The corresponding scale of reference used by the Planck mission is $k_{\ast} = 0.05\, \textrm{Mpc}^{-1}$, and its amplitude is $\ln(10^{10}A_{s}) = 3.089\, \pm\, 0.036\, (68\%\, \textrm{CL})$ \cite{planck-inf}. As an additional constraint, we must check that the obtained pivot scale is indeed a plausible scale, observed nowadays in the CMB. This imposes a condition in the number of e-folds $N_\star$ that took place from the exit of the pivot scale beyond the Hubble horizon until the end of inflation. Explicitly, $N_\star = \ln(a_{\rm end}/a_\star)$, where $a_\star$ and $a_{\rm end}$ are the values of the scale factor at the time when the pivot scale crossed the horizon and when inflation ended, respectively. Taking into account the bounds given in Ref. \cite{bib:LiddleLeach2003} and that we are considering the pivot scale $k_{\ast}$ (and not the horizon scale at present, $k_0 = a_{0} H_{0} \approx k_{\ast}/220$), we have checked that $65 > N_\star > 45$. Let us remark, nevertheless, that this range of values depends considerably on the behavior and duration of the reheating phase, as well as on the transition to it from the inflationary phase (see Refs. \cite{bib:MartinRingeval2010,encyc} for a better treatment of this reheating phase). Table \ref{table:kpivot} gives the corresponding value of the pivot scale that results from our scale-matching method for different choices of the homogeneous initial conditions. There, we also list the value of several background quantities at the time of the horizon crossing, as well as the (local) spectral index for the pivot scale.
\def\arraystretch{1.2}
\begin{table}[h]
\begin{center}
\begin{tabular}{|c|c|c|c|c|c|c|c|}
\hline
$\phi_B$ & $m\, (\cdot 10^{-6})$ & $k_\star$ & $N_\star$ & $n_s$ & $\phi_\star$ & $ V_\star \,(\cdot 10^{-12})$ & $H_\star \,(\cdot 10^{-6})$  \\
\hline
$ 0{.}950$ & $1{.}180$  & $0{.}03013$ & $61{.}0226$ & $0{.}96694$ & $3{.}10229 $ & $6{.}70038$ & $7{.}50248$ \\
$ 0{.}960$ & $1{.}180$  & $0{.}04415$ & $61{.}0228$ & $0{.}96704$ & $3{.}10230 $ & $6{.}70041$ & $7{.}50250$ \\
$ 0{.}970$ & $1{.}180$  & $0{.}06486$ & $61{.}0220$ & $0{.}96704$ & $3{.}10228 $ & $6{.}70031$ & $7{.}50245$ \\
$ 0{.}980$ & $1{.}180$  & $0{.}09528$ & $61{.}0224$ & $0{.}96699$ & $3{.}10229 $ & $6{.}70035$ & $7{.}50248$ \\
$ 0{.}990$ & $1{.}180$  & $0{.}13996$ & $61{.}0238$ & $0{.}96701$ & $3{.}10232 $ & $6{.}70052$ & $7{.}50256$ \\
\hdashline
$ 0{.}950$ & $1{.}190$  & $0{.}04786$ & $60{.}5087$ & $0{.}96681$ & $3{.}08912 $ & $6{.}75668$ & $7{.}53402$ \\
$ 0{.}960$ & $1{.}190$  & $0{.}07015$ & $60{.}5087$ & $0{.}96672$ & $3{.}08912 $ & $6{.}75669$ & $7{.}53403$ \\
$ 0{.}970$ & $1{.}190$  & $0{.}10280$ & $60{.}5100$ & $0{.}96673$ & $3{.}08915 $ & $6{.}75683$ & $7{.}53411$ \\
$ 0{.}980$ & $1{.}190$  & $0{.}15101$ & $60{.}5102$ & $0{.}96673$ & $3{.}08916 $ & $6{.}75685$ & $7{.}53412$ \\
$ 0{.}990$ & $1{.}190$  & $0{.}22233$ & $60{.}5093$ & $0{.}96673$ & $3{.}08913 $ & $6{.}75674$ & $7{.}53406$ \\
\hdashline
$ 0{.}950$ & $1{.}200$  & $0{.}07534$ & $60{.}0045$ & $0{.}96646$ & $3{.}07614 $ & $6{.}81309$ & $7{.}56550$ \\
$ 0{.}960$ & $1{.}200$  & $0{.}11041$ & $60{.}0044$ & $0{.}96646$ & $3{.}07614 $ & $6{.}81308$ & $7{.}56549$ \\
$ 0{.}970$ & $1{.}200$  & $0{.}16218$ & $60{.}0032$ & $0{.}96644$ & $3{.}07610 $ & $6{.}81294$ & $7{.}56542$ \\
$ 0{.}980$ & $1{.}200$  & $0{.}23823$ & $60{.}0031$ & $0{.}96644$ & $3{.}07610 $ & $6{.}81294$ & $7{.}56542$ \\
$ 0{.}990$ & $1{.}200$  & $0{.}34995$ & $60{.}0044$ & $0{.}96645$ & $3{.}07613 $ & $6{.}81308$ & $7{.}56549$ \\
\hdashline
$ 0{.}950$ & $1{.}210$  & $0{.}11776$ & $59{.}5075$ & $0{.}96616$ & $3{.}06329 $ & $6{.}86938$ & $7{.}59678$ \\
$ 0{.}960$ & $1{.}210$  & $0{.}17258$ & $59{.}5073$ & $0{.}96617$ & $3{.}06329 $ & $6{.}86935$ & $7{.}59676$ \\
$ 0{..}970$ & $1{.}210$  & $0{.}25293$ & $59{.}5083$ & $0{.}96617$ & $3{.}06331 $ & $6{.}86946$ & $7{.}59682$ \\
$ 0{.}980$ & $1{.}210$  & $0{.}37154$ & $59{.}5080$ & $0{.}96617$ & $3{.}06331 $ & $6{.}86944$ & $7{.}59681$ \\
$ 0{.}990$ & $1{.}210$  & $0{.}54702$ & $59{.}5068$ & $0{.}96616$ & $3{.}06327 $ & $6{.}86930$ & $7{.}59673$ \\
\hdashline
$ 0{.}950$ & $1{.}220$  & $0{.}18281$ & $59{.}0180$ & $0{.}96589$ & $3{.}05058 $ & $6{.}92557$ & $7{.}62787$ \\
$ 0{.}960$ & $1{.}220$  & $0{.}26730$ & $59{.}0199$ & $0{.}96588$ & $3{.}05063 $ & $6{.}92580$ & $7{.}62800$ \\
$ 0{.}970$ & $1{.}220$  & $0{.}39264$ & $59{.}0184$ & $0{.}96588$ & $3{.}05059 $ & $6{.}92562$ & $7{.}62790$ \\
$ 0{.}980$ & $1{.}220$  & $0{.}57677$ & $59{.}0180$ & $0{.}96588$ & $3{.}05059 $ & $6{.}92558$ & $7{.}62788$ \\
$ 0{.}990$ & $1{.}220$  & $0{.}84723$ & $59{.}0190$ & $0{.}96688$ & $3{.}05061 $ & $6{.}92568$ & $7{.}62793$ \\
\hline
\end{tabular}
\caption{For each pair of parameters $(\phi_B, m)$, this table provides the corresponding pivot scale, the number of e-folds until the end of inflation, the (local) spectral index, the value of the scalar field, the value of the potential, and the Hubble parameter. All background quantities are evaluated at the time when the pivot scale crosses the Hubble horizon.}
\label{table:kpivot}
\end{center}
\end{table}

Let us also comment briefly on the issue of the weak gravitational lensing corrections. This phenomenon affects the trajectory of the photons in the CMB from the last scattering surface until today. This lensing is caused by the gravitational potential of large scale structures \cite{lens1}, and therefore it is ultimately determined by the power spectrum of the cosmological perturbations. In this work, we follow the method of Ref. \cite{lensing}, which gives accurate results for all scales, provided that the non-Gaussianities that are due to nonlinear effects are not important. In absence of lensing, the only source of power in the $BB$-correlation function are the tensor perturbations \cite{b-tensor}. However, the gravitational lensing produces a mixing between $E$ and $B$ polarizations, mixing that is not negligible at small scales. 

\begin{figure}
\centering     
\includegraphics[width = 0.49\textwidth]{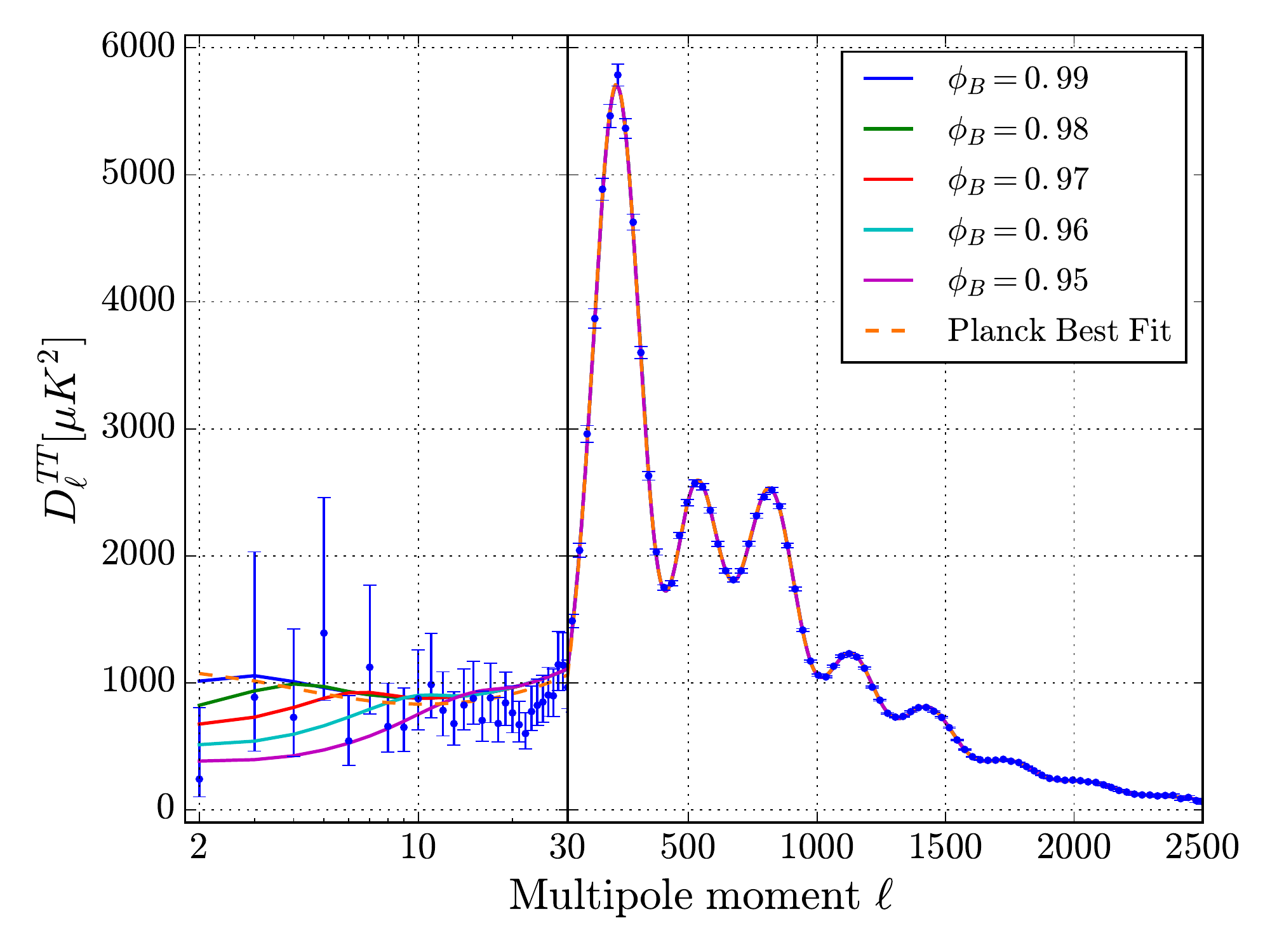}   
\includegraphics[width = 0.49\textwidth]{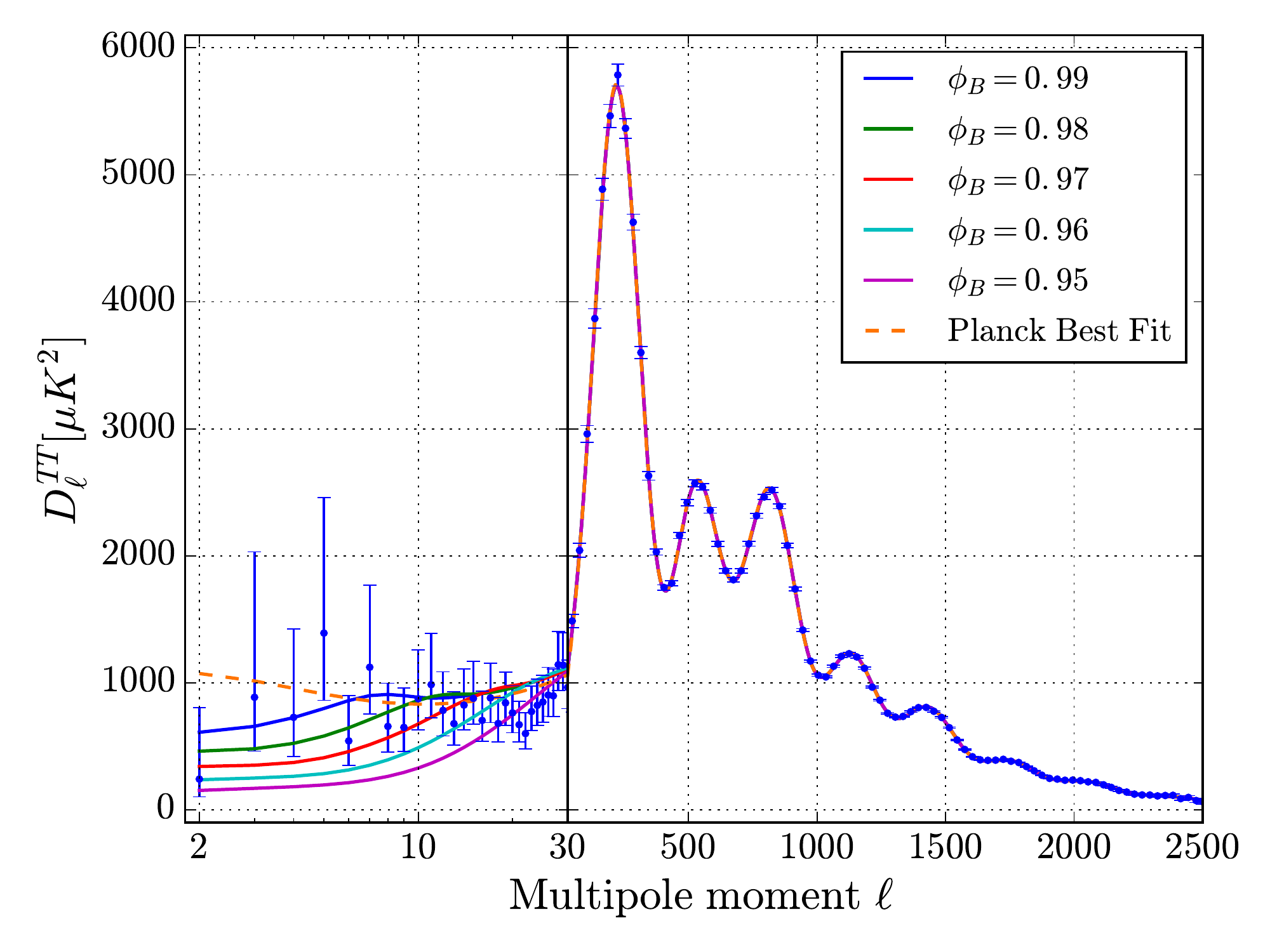}
\caption{$TT$ angular power spectrum provided by Planck best fit and spectra computed for the non-oscillating vacuum with different values of the scalar field at the bounce. Left panel: $m=1.20\cdot10^{-6}$. Right panel: $m=1.18\cdot10^{-6}$. The corresponding cosmological parameters determined by Planck best fit for TT+lowP data are given in Ref. \cite{planck-inf}. Both panels incorporate lensing corrections.}   
\label{fig:TTanist}
\end{figure}

To account for this lensing, we have computed again (see also Ref. \cite{hybr-pred}) the spectrum of the temperature anisotropies of the CMB, but this time including also tensor perturbations in our computations, and assuming that the perturbations are in the non-oscillating vacuum at the bounce. In these computations, we have employed the \textsf{CLASS} code \cite{class}. We have considered the base $\Lambda$CDM model with the best-fit values of the baryon density, cold dark matter density, angular size of the sound horizon, and Thompson scattering optical depth for the TT+lowP data. These best-fit values are given in the first column of Table 4 in Ref. \cite{planck}. The results are summarized in Fig. \ref{fig:TTanist}. The addition of tensor perturbations produces an increase of power at low multipole moments $\ell$. However, if the modes that cross the Hubble horizon today are within the suppression region that characterizes the non-oscillating vacuum state, this enhancement in the spectrum turns out to affect modes with $\ell\sim20$. The enhancement is very small, in any case. Besides, although we have not carried out a rigorous statistical analysis, it seems that the calculations that include lensing are in better agreement with the observational data, as well as with the best fit of the Planck Collaboration. The lensing effect modifies the spectrum of the anisotropies mainly at large multipole moments. Roughly speaking, its contribution reduces the amplitude of the oscillations of the baryonic resonances.

\begin{figure}
\centering     
\includegraphics[width = 0.49\textwidth]{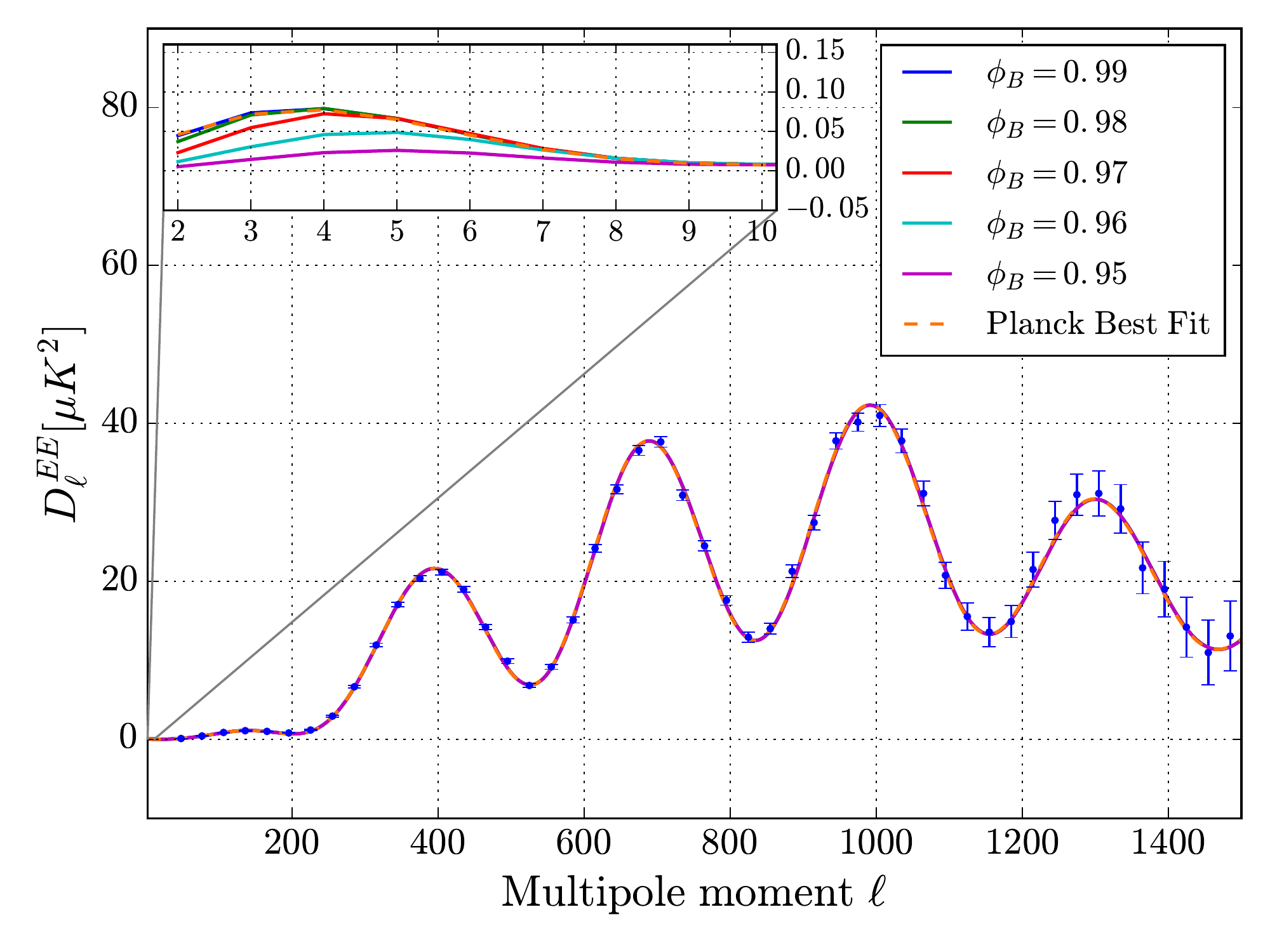}  
\includegraphics[width = 0.49\textwidth]{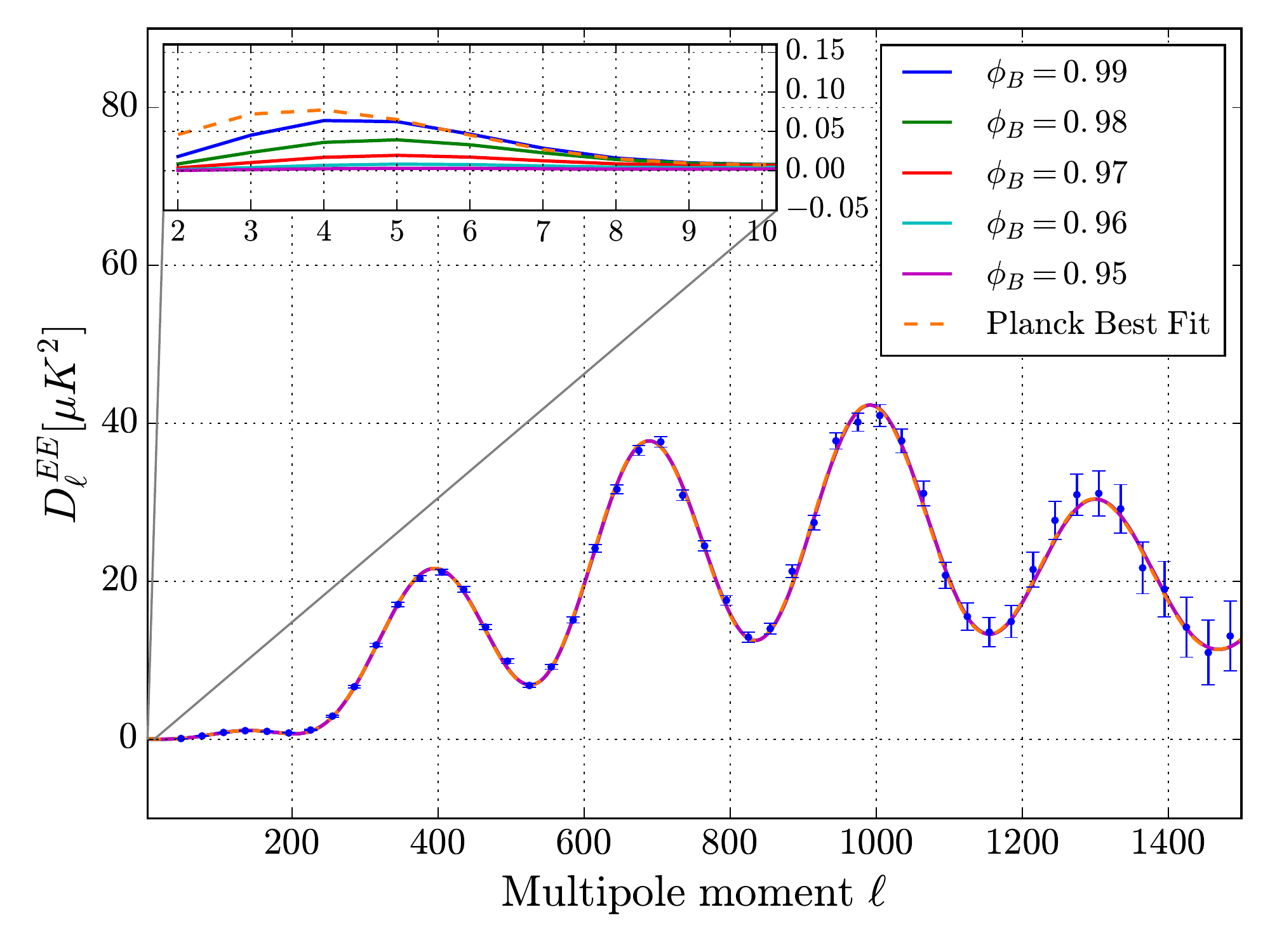}\\
\caption{$EE$ angular power spectrum provided by Planck best fit and spectra computed for the non-oscillating vacuum with different values of the scalar field at the bounce. Left panel: $m=1.20\cdot10^{-6}$. Right panel: $m=1.18\cdot10^{-6}$. The corresponding cosmological parameters determined by Planck best fit for TT+lowP data are given in Ref. \cite{planck-inf}. Both panels incorporate lensing corrections.}  
\label{fig:EEanist}
\end{figure}

We have also computed the spectrum of other correlation functions and compared them with observational data. For instance, in Fig. \ref{fig:EEanist} we display the $EE$-correlation function. We see that the suppression of the primordial power spectrum reduces the amplitude of this correlation function at very low multipole moments $\ell$, in comparison with the behavior of the Planck best fit. The lensing, as in the $TT$-correlation function, gives a better fit to the observations at large $\ell$.

\begin{figure}
\centering     
\includegraphics[width = 0.49\textwidth]{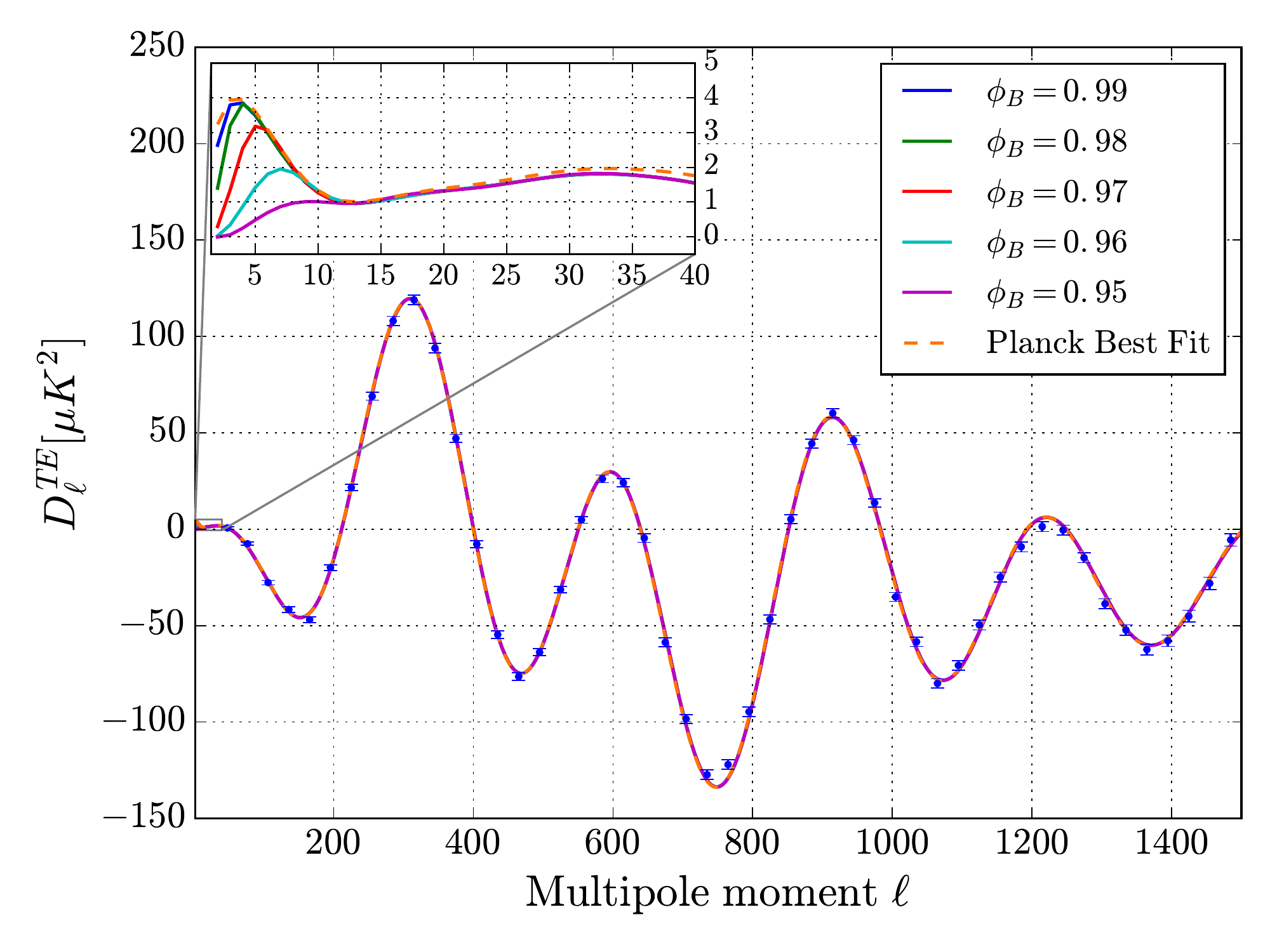}  
\includegraphics[width = 0.49\textwidth]{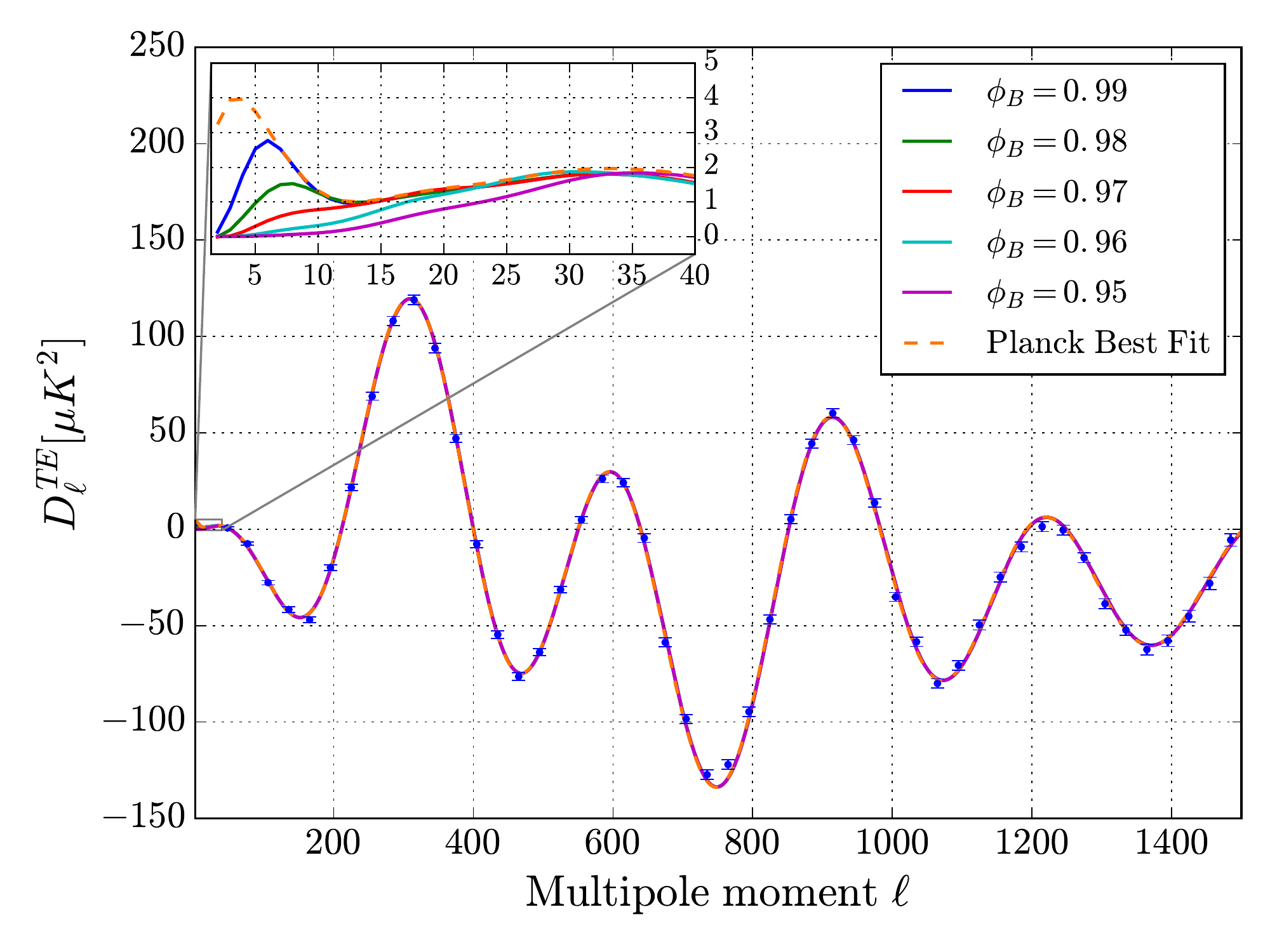}\\
\caption{$TE$ angular power spectrum provided by Planck best fit and spectra computed for the non-oscillating vacuum with different values of the scalar field at the bounce. Left panel: $m=1.20\cdot10^{-6}$. Right panel: $m=1.18\cdot10^{-6}$. The corresponding cosmological parameters determined by Planck best fit for TT+lowP data are given in Ref. \cite{planck-inf}. Both panels incorporate lensing corrections.}   
\label{fig:TEanist}
\end{figure}

In addition, in Fig. \ref{fig:TEanist} we plot the angular power spectrum for the $TE$-cross-correlation function. Again, the behavior of the best fit provided by the Planck Collaboration at low multipole moments differs from the predictions obtained here for the non-oscillating vacuum when the suppression of the primordial power spectrum becomes relevant in the modes that cross the Hubble horizon today. We also notice that, at large $\ell$, the lensing plays again an important role in improving the agreement between predictions and observations.

\begin{figure}
\centering     
\includegraphics[width = 0.49\textwidth]{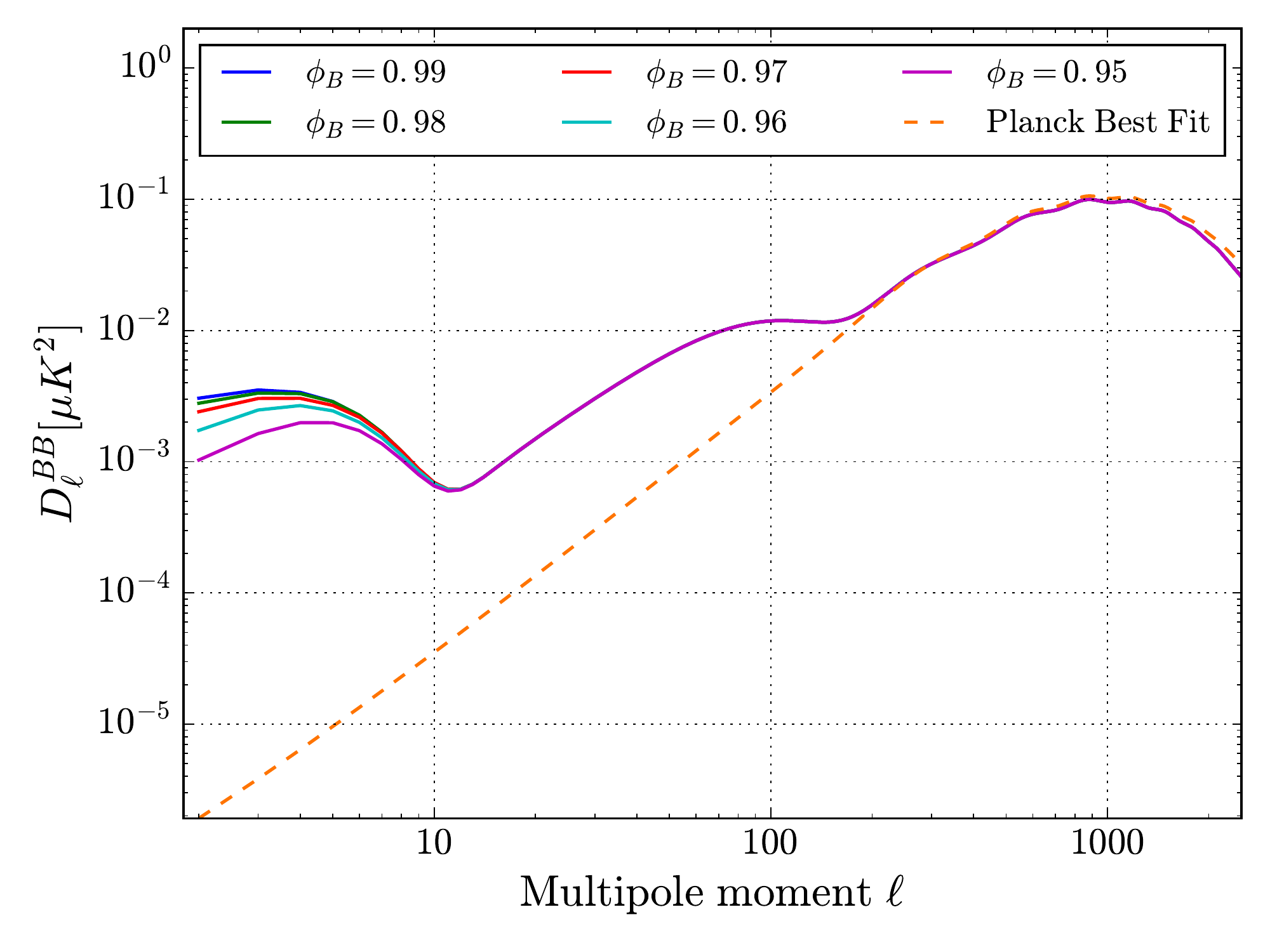}  
\includegraphics[width = 0.49\textwidth]{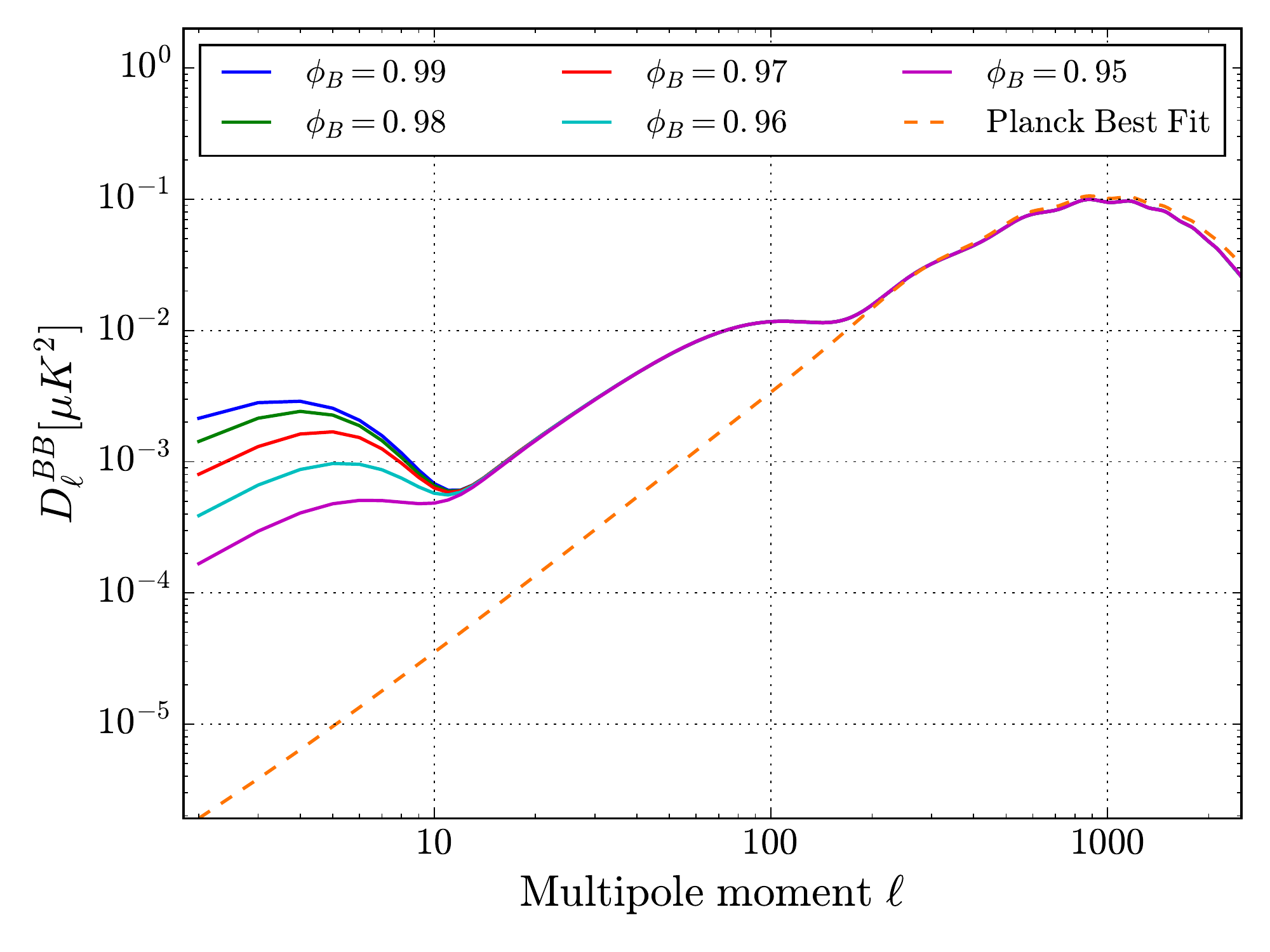}
\caption{$BB$ angular power spectrum provided by Planck best fit and spectra computed for the non-oscillating vacuum with different values of the scalar field at the bounce. Left panel: $m=1.20\cdot10^{-6}$. Right panel: $m=1.18\cdot10^{-6}$. The corresponding cosmological parameters determined by Planck best fit for TT+lowP data are given in Ref. \cite{planck-inf}. Both panels incorporate lensing corrections.}   
\label{fig:BBanist}
\end{figure}

Finally, we have compared the $BB$-correlation function of the non-oscillating vacuum with the theoretical value predicted for it by the Planck Collaboration. This comparison can be found in Fig. \ref{fig:BBanist}. We observe a good agreement between the spectra at large multipole moments $\ell$, but important differences arise at low $\ell$. One of the reasons is the fact that the predictions reached by Planck ignore tensor perturbations in the calculations. For large $\ell$, the main contribution to the power spectrum comes from weak gravitational lensing, which is known to be responsible of the large amplitude in that region (see, for instance, Ref. \cite{lensing}). However, lensing does not contribute significantly to the spectrum at low $\ell$. Indeed, the BICEP Collaboration \cite{bicep} studied some few years ago the $BB$-correlation function in the interval $30\lessapprox\ell\lessapprox150$. This interval is actually in the region of multipole moments that is not considerably contaminated by lensing. Consequently, the presence of power in this region, had it not been explained eventually by other sources, would have been a strong evidence of the presence of primordial tensor modes in the CMB. 

In summary, if the suppression of power in the primordial power spectrum of the non-oscillating vacuum is relevant for modes that are in the large scale sector today, such effect would translate into a decrease of power in the correlation functions at small multipole moments. This is the main characteristic of the non-oscillating vacuum within our hybrid approach: the suppression of power at low $\ell$ in the studied correlation functions of the CMB. 

\section{Discussion and conclusions}\label{sec:dis-conc}

In this work, we have discussed possible physical consequences of the hybrid quantization approach in LQC \cite{hybr-inf1,hybr-inf2,hybr-inf3,hybr-inf4,hybr-pred,hybr-ten} on the behavior of cosmological perturbations in an inflationary universe. More specifically, we have considered a flat FRW spacetime coupled to a massive scalar field. In this system, scalar and tensor perturbations have been introduced in order to account for the small inhomogeneities that originated the large scale structures of our Universe. We have mostly focused on the analysis of tensor perturbations, because the scalar ones were already studied and (at least partially) compared with observations in Ref. \cite{hybr-pred}. We emphasize that, in both cases (i.e., for scalar and tensor perturbations), the backreaction has been ignored in the discussion, treating the perturbations as as test fields. Besides, we have considered that the wave function of the system can be factorized as in Eq. \eqref{BOans}, namely, separating the dependence on the background geometry from that on the perturbations. This kind of Born-Oppenheimer ansatz allows us to deal with the evolution of the FRW geometry independently of the inhomogeneities. In addition, we have concentrated our analysis on quantum states of the background geometry that are sufficiently and suitably peaked, so that the effective dynamics of LQC is valid to describe the evolution of the peak trajectory and any relevant expectation value associated with it. We have also admitted the reasonable hypothesis that the quantum dependence on the perturbations, with a Hamiltonian that is quadratic, has a direct effective counterpart in which creation and annihilation operators are replaced with classical variables. Finally, we have followed a standard treatment of those perturbations subject to such an effective dynamics.

In this way, we have deduced effective equations of motion for the perturbations (see Ref. \cite{hybr-ten}). They take the form of an infinite collection of decoupled ordinary differential equations with a time-dependent mass. These equations can be easily integrated provided that suitable initial data are given. In Fig. \ref{fig:tdmass}, we compare the time-dependent masses of the scalar and the tensor perturbations in the hybrid and the dressed metric \cite{AAN2} approaches (in both cases, once backreaction is ignored and the effective description is accepted). These time-dependent masses agree with their values in general relativity away from the bounce. Moreover, the values of the mass for the scalar and the tensor perturbations esentially coincide in each of the two approaches separately, at least for background solutions that are kinetically dominated at the bounce. But, in general, these values differ in the two approaches, especially in regimes where the quantum corrections are important, as it actually happens around the bounce, so that they can even get opposite signs. In conclusion, although both approaches provide effective equations for the perturbations that are similar and share several qualitative aspects, the way in which they incorporate quantum gravity corrections in those equations leads to significant differences.

Contrary to the typical situation in standard general relativity, in LQC there seems to exist a privileged Cauchy surface where initial data can be supplied: the quantum bounce. Nevertheless, since quantum gravity corrections are important at this bounce, it is not completely clear which choice of initial data should be adopted for the scalar and the tensor perturbations there. At least, it is natural to assume that both types of perturbations start in the same vacuum state. Actually, there exist several prescriptions in the literature to determine such initial data. For instance, adiabatic states seem an appealing choice for very large wavenumbers, because they are approximate solutions with a convenient splitting between positive and negative frequency contributions in the ultraviolet sector, roughly speaking. However, the adiabatic approximation breaks down for solutions corresponding to large scale modes. These modes can be subject to strong curvature effects during their evolution and experience excitations that can produce an enhancement in the power spectrum. In these circumstances, the adiabatic approximation is not appropriate. In fact, this is a common problem shared by the hybrid and the dressed metric approaches for the treatment of cosmological perturbations in LQC. With this motivation in mind, a new vacuum was proposed for the perturbations in Ref. \cite{hybr-pred}: the so-called non-oscillating vacuum state. This state minimizes the particle creation along the evolution, and in this sense can be regarded as the best adapted to the background. In particular, this affects the splitting between positive and negative frequency solutions, even for some large scale modes (at least for those that oscillate a sufficient number of times between the bounce and the onset of inflation). For the scalar perturbations, this vacuum has an associated power spectrum that agrees with that of standard slow-roll inflation, even for those scales that notice the large value of the curvature of the spacetime in the quantum regime. Besides, it is remarkable that this vacuum shows a suppression of power at large scales (low multipole moments), in agreement with what is apparently suggested by the present observations of the anisotropies of the CMB.

In our study, we have investigated in detail the evolution of the tensor perturbations in LQC for the hybrid quantization approach (without backreaction and within the effective approximation), assuming that their initial state at the bounce is an adiabatic vacuum state of 0th, 2nd, or 4th order. In addition, we have also considered the possibility of a non-oscillating vacuum state, as proposed in Ref. \cite{hybr-pred}. We have computed the power spectrum of the tensor perturbations for all of these vacua. In the case of the adiabatic states, we have observed a strong particle production at scales that feel the curvature of the background in the quantum regime. One of the main conclusions is that, in LQC, the existence of a bounce introduces an upper bound on the possible enhancement of the power spectrum (by particle creation), a fact which is in clear contrast with the situation found in general relativity when one approaches the cosmological singularity. This seems a robust prediction in LQC, valid for both the hybrid and the dressed metric approaches. Another important result of our study is a surprising property of the non-oscillating vacuum: it behaves asymptotically as a high-order adiabatic state. In the left panel of Fig. \ref{fig:beta} we display the absolute value of the antilinear coefficients $\beta_{k}$ (multiplied by $k^{3/2}$) of the Bogoliubov transformation between this vacuum and several adiabatic states. We see that, the higher the adiabatic order is, the faster the decay is for large $k$. In the right panel, we observe a similar behavior when we consider transformations between adiabatic states. The rate of convergence is determined by the adiabatic state of lower order. From our analysis, we conclude that the non-oscillating vacuum belongs in fact to the equivalence class of the adiabatic vacua, with the same asymptotic behavior as an adiabatic state of at least 4th order.

We have also computed the tensor-to-scalar ratio derived from our hybrid quantization approach. With the family of vacuum states that we have considered here, and the results of Ref. \cite{AAN2}, the ratio appears to be constant, even at scales where the quantum gravity corrections are important (provided that one gets rid of the strong oscillations). This is our next important conclusion: the tensor-to-scalar ratio seems approximately constant in LQC, regardless of the initial state and the adopted quantization approach, when one assumes that the scalar and the tensor perturbations have the same vacuum. One might have guessed this result, given our assumption about the coincidence of the vacua and the fact that the time-dependent masses of both the scalar and the tensor perturbations are very similar everywhere (for background spacetimes that are kinetically dominated at the bounce). On the other hand, if one takes into account the strong oscillations, the tensor-to-scalar ratio also oscillates around a constant value. These oscillations are caused by the difference between the phases of the scalar and the tensor modes at the horizon crossing. Besides, we observe that the consistency relation between the tensor-to-scalar ratio and the tensor spectral index is satisfied for sufficiently large values of $k$. However, the relation is violated for adiabatic states that entail an important enhancement of the power, in the region of wavenumbers $k$ where this power increase occurs. Most remarkably, nonetheless, the non-oscillating vacuum turns out to be compatible with the consistency relation up to very small wavenumbers $k$.  

Finally, we have computed the $TT$, $EE$, $TE$, and $BB$ correlation functions for the non-oscillating vacuum. For this purpose, we have employed the \textsf{CLASS} code. We have compared our predictions with the best fit of the TT+lowP data of the Planck Collaboration, assuming a base $\Lambda$CDM model with value of the cosmological parameters given in the first column of Table 4 in Ref. \cite{planck}. To get a better fit of the spectrum, we have introduced corrections caused by cosmological lensing. These corrections introduce slight modifications of the amplitude of the baryonic peaks, improving the fit at small scales. In addition, the $BB$-correlation function is considerably affected by these corrections at relatively large multipole moments. Our numerical computations indicate that a general property of the non-oscillating vacuum is the suppression of power at low multipole moments $\ell$ (large scales), an effect that we have noticed in all the studied correlation functions. Therefore, a suitable choice of vacuum state, based on first principles, might suffice to explain the plausible lack of power suggested by present observations. 

In summary, LQC provides a powerful formalism for the study of cosmological perturbations in inflation that leads to robust predictions, even though some phenomena crucially depend on the choice of initial state for the perturbations and on the concrete quantization approach. Robust predictions of this type are the existence of a bound on the particle production and the intrinsic similarities between the dynamics of the scalar and the tensor perturbations (at least for kinetically dominated bounces). Furthermore, these predictions are in good agreement with the available observations. In this way, LQC is able to connect successfully the physics of the early Universe at the Planck regime with the present observations, extending the traditional formalism based in general relativity.

\acknowledgments

The authors are thankful to M. Mart\'{\i}n-Benito for discussions and enlightening comments, and to T. Paw{\l}owski for allowing them to use his computational equipment. This work was supported by the Project. No. MINECO FIS2014-54800-C2-2-P from Spain. D. M-dB acknowledges financial support from the Project No. CONICYT/FONDECYT/POST\-DOCTORADO/3140409 from Chile. J.O. was supported by the Project No. NSF-PHY-1305000, the Project No. PHY-1505411, the Eberly research funds of Penn State University (USA), and by Pedeciba (Uruguay).

\end{document}